\def\be{\begin{equation}}
	\def\ee{\end{equation}}
\def\ba{\begin{array}}
	\def\ea{\end{array}}

\def\mathbi#1{\text{\em #1}}

\documentclass[prl,showpacs,twocolumn,amsmath]{revtex4}
\usepackage{amsmath}
\usepackage{amsfonts}
\usepackage{mathrsfs}
\usepackage{amssymb}
\usepackage{pifont}
\usepackage{epsfig,subfigure,dsfont,amsthm,amsbsy,mathrsfs,amscd}
\usepackage{epstopdf}
\usepackage{bbm}
\usepackage{color}
\def\qed{\leavevmode\unskip\penalty9999 \hbox{}\nobreak\hfill
	\quad\hbox{\leavevmode  \hbox to.77778em{%
			\hfil\vrule   \vbox to.675em%
			{\hrule width.6em\vfil\hrule}\vrule\hfil}}
	\par\vskip3pt}
\usepackage{leftidx}

\input amssym.def
\begin{document}
	\title{\large\bf The evolution of quantum battery capacity of GHZ-like states under Markovian channels}
	\author{Hui Liu, Tinggui Zhang$^{\dag}$}
	\affiliation{ School of Mathematics and Statistics, Hainan Normal University, Haikou, 571158, China \\
		$^{\dag}$ Correspondence to tinggui333@163.com}

	\bigskip
	\bigskip
	\begin{abstract}Quantum battery has enormous potential for development, and quantum battery capacity is an important indicator of quantum battery. In this work, we mainly study the evolution of quantum battery capacity of GHZ state and GHZ-like states under Markovian channels in the tripartite system. We find that under the depolarizing channel and bit-phase flip channel, the battery capacity shows a brief sudden death of the capacity. And we also find that under the dephasing channel, the battery capacity gradually decreases and tends to a constant, that is, the frozen capacity. We show that the battery capacity monotonically decreases for GHZ state under the amplitude damping channel on the first subsystem. And we study the variation of capacity under the Markovian channels n times on the first subsystem using the GHZ state. We can observe that under the amplitude damping and dephasing channels, the battery capacity decreases and tends to a constant, i.e. frozen capacity, and the larger n, the earlier this phenomenon occurs. We also investigate the evolution of capacity under three independent same type Markovian channels. We have also conducted corresponding research on GHZ-like states.
		
	\end{abstract}
	
	 \maketitle
	
	\section{I. Introduction}
Quantum battery is a novel energy storage device that utilizes quantum mechanical properties such as quantum coherence, entanglement, and collective effects to store and release energy \cite{sqxw,mkmn,mlap,gmam,jmba,tagk}. Compared to traditional batteries, quantum batteries theoretically have the potential for ultra-fast charging, high power output, and nearly lossless energy transfer, making them important applications in micro quantum devices, quantum computing, and future energy networks. The concept of quantum batteries originates from interdisciplinary research in quantum thermodynamics and quantum information science. Alicki and Fannes formally introduced the concept of quantum batteries in the context of information theory, characterized by the maximum energy that could be extracted from a quantum system through characterization under unitary operations\cite{ramf}. Ferraro et al. proposed the collective charging mechanism in Ref.\cite{dmgv}, demonstrating that quantum entanglement can improve charging efficiency. In Ref.\cite{fffl}, Campaioli et al. quantified the power upper limit of quantum batteries and demonstrated that they were limited by the uncertainty relationship of quantum mechanics, but still far higher than classical batteries, which has driven optimization design. So far, quantum batteries have been widely studied both theoretically and experimentally\cite{ramf,dmgv,fffl,jcaa,gfrz,dgam,mmrf,ra,dgmp,dgdm,stam,fbla,sbad,jddr,jjcf,fmaj,ptzg,mgmk,jpjt,tfmf,srsv,dxhp,admm,brms,rgvg,srsg,
mham,fqfm,dmsm,xxdw,pgvf,gfld,xyxj,ths,paku,arbp,asmt,hkxh,fmfq,rcgb,yxt,xymr,rbgp,ks,fsjm,ylts,mail,yhst,aaka,mbsg,svak,zgxc,szzz,
yyxs,lsfh}. Many theoretical protocols related to quantum batteries have also been proposed, such as using the collective behaviour of quantum systems to achieve super-linear charging rates\cite{dgam,dgdm,jddr,hkxh,xxdw,jjcf,pgvf,fmaj,ptzg}, using the non classical coherence of quantum states to improve energy storage efficiency\cite{gfld,mmrf,mgmk,fbla,aaka}, a quantum battery was constructed based on spin chain models such as the Heisenberg model\cite{mbsg,svak}. And the problems of actively extracting energy through quantum measurements such as weak measurements and projective measurements were studied\cite{ths,paku,arbp,yhst}. More research can be seen in the review of quantum batteries \cite{fsjm}.
	
Markovian channel is an important model in quantum information science that describes the dynamics of open quantum systems. Its core feature is that the system evolution depends only on the current state and satisfies quantum Markovian property (memorylessness).  The Markovian channel, as the core model of open quantum system theory, provides an important framework for studying the dynamic evolution of quantum battery capacity. In Ref.\cite{ylts}, the authors systematically explored the capacity evolution behavior of Bell-diagonal states under Markovian channels, including bit flip channel, depolarizing channel and amplitude damping channel. They also discovered two special phenomena: capacity sudden death and capacity freezing. In this study, we investigate the evolution of the quantum battery capacity of GHZ states and GHZ-like states under Markovian channels. We find that the three-qubit GHZ state retains a certain level of energy storage capability under the amplitude damping channel, while the GHZ-like states exhibits better stability in energy storage under the dephasing channel. More importantly, the initial state remains at a relatively high capacity level even after undergoing multiple quantum noise channel effects, which may contribute to the design of many-body quantum batteries.
	
	The rest of this paper is organized as follows. In the second section, we study the evolution of quantum battery capacity for GHZ state under the Markovian channels once and n times. In the third section, we explore the evolution of quantum battery capacity under the Markovian channels once and n times using the GHZ-like states. We present a summary in the last section.

\section{II. Quantum battery capacity of GHZ state in tripartite system under Markovian channel}
A critical metric for assessing the performance of quantum batteries is the capacity of the quantum battery\cite{stam,asmt,xymr,yxt}. Unlike classical batteries, the quantum batteries capacity depends not only on the physical properties of the materials themselves, but also on quantum coherence \cite{ks}, entanglement, and collective effects. In Ref.\cite{srsv}, the authors proposed an important quantitative indicator related to quantum batteries, i.e. the ergotropy functional,
$$
\varepsilon(\rho,H)=\max_{U\in \mathbb{U}(d)}\{Tr(\rho H)-Tr(U \rho U^{\dagger}H)\},
$$
where $\rho$ is the quantum state of d-dimensional Hilbert space $\mathcal{H}$ with Hamiltonian $H$ and $\mathbb{U}(d)$ represents the set of unitary operations acting on $\mathcal{H}$ and the maximum takes over the states under all unitary evolution $\rho \rightarrow U \rho U^{\dagger}$.
	Recently, a new definition of quantum battery proposed by Yang et al. has emerged. This new definition in Ref.\cite{xymr} is quantum battery capacity, which is an important indicator of quantum batteries,
	\begin{equation}\label{e1}
		\begin{split}
			\mathcal{C}(\rho;H)&=\sum_{i=0}^{d-1}\epsilon_i(\lambda_i-\lambda_{d-1-i})\\
			&=\sum_{i=0}^{d-1}\lambda_i(\epsilon_i-\epsilon_{d-1-i}),
		\end{split}
	\end{equation}
	where $\{\lambda_i\}$ represents eigenvalues of quantum state and satisfies $\lambda_0\leqslant \lambda_1\leqslant...\leqslant \lambda_{d-1-i}$, and $\{\epsilon_i\}$ represents energy levels of Hamiltonian $H=\sum_{i=0}^{d-1}\epsilon_i|\varepsilon_i \rangle \langle \varepsilon_i|$ and satisfies $\epsilon_0\leqslant \epsilon_1 \leqslant...\leqslant \epsilon_{d-1-i}$. $\mathcal{C}(\rho;H)$ serves as a critical indicator for assessing the quality of a quantum battery, establishing the theoretical upper limit for energy extraction through unitary evolution within the quantum framework. Besides, quantum battery capacity $C(\rho,H)$ is a convex and sub-linear functional.
	\\
	In Ref.\cite{ylts}, authors studied the evolution of the quantum battery capacity for the Bell-diagonal state under Markovian channels and studied the dynamics of the quantum battery capacity of the Bell-diagonal state under two independent same type local Markovian channels. In this work, we mainly study the quantum battery capacity evolution of GHZ state and GHZ-like states in the tripartite system under Markovian channels. The concept of Markovian channels is applied to the research of quantum batteries, primarily to analyze and optimize their performance in real physical environments, particularly regarding capacity and charging efficiency. Markovian channels offer a rigorous mathematical framework for describing and quantifying the impact of environmental noise on the quantum properties of batteries, which in turn limits their extractable energy and charging power. As the cornerstone of open quantum system dynamics, Markovian channels theory serves as an essential bridge connecting the ideal quantum battery theory to its practical applications.
	
	Let us consider the three-qubit GHZ state,
	\begin{equation}\label{e2}
			\rho_{ABC}=|GHZ\rangle \langle GHZ|
	\end{equation}
	where $|GHZ\rangle =\frac{1}{\sqrt 2}(|000\rangle +|111\rangle)$. The Hamiltonian of the tripartite system is $ H_{ABC}=\epsilon^A \sigma_3 \otimes I_2 \otimes I_2+ \epsilon^BI_2 \otimes \sigma_3 \otimes I_2+ \epsilon^C I_2 \otimes I_2 \otimes \sigma_3$, where $\sigma_3$ is the standard Pauli matrices, $I_2$ is the identity operator, and $0 \leqslant \epsilon^C \leqslant \epsilon^B \leqslant \epsilon^A $. The detailed calculation process can be found in the Appendix A. Therefore, from (\ref{e1}) we have
	
	{\begin{equation}\label{e3}
			\mathbi{C}(\rho_{ABC},H_{ABC})=2(\epsilon^A+\epsilon^B+\epsilon^c).
	\end{equation}
	The reduced state $ \rho_{ABC} $ are $ \rho_A=\frac{I}{2}=\rho_B=\rho_C$. Accordingly, we have $ \mathbi{C}(\rho_A,H_A)=0=\mathbi{C}(\rho_B,H_B)=\mathbi{C}(\rho_C,H_C) $.
\subsection{A. Quantum battery capacity evolution of GHZ state under Markovian channels on the first subsystem}
	In this section, we mainly investigate the evolution of quantum state $\rho$ under a trace-preserving quantum operation $\varepsilon(\rho)$ \cite{mail},
	\begin{equation}\label{e4}
	\varepsilon(\rho)=\sum_{i,j,k}(E_i \otimes E_j \otimes E_k)\rho(E_i \otimes E_j \otimes E_k)^{\dagger},
    \end{equation}   
	where $\{E_m\}$ is the set of Kraus operators and satisfies completeness relationship, i.e $\sum_{m}E_m^{\dagger}E_m=I$. The  trace-preserving quantum operation describes the evolution of a tripartite quantum system subjected to local, independent, and parallel quantum noise processes. This noise does not induce any correlations between the three subsystems; rather, the noise process for each subsystem is solely determined by its corresponding set of Kraus operators. The principle of traceability guarantees that the entire evolution process adheres to the law of conservation of probability. Table I \cite{yszj} lists several kraus operators of typical channels, where phase flip channel (pf) and dephasing channel (dp) preserve the form of the density operator $\rho_{ABC}$.
	\\ 
	\begin{table*}[htbp] 
		\centering
		\begin{tabular}{|c|c|}
			\hline
			channel&Kraus operators\\
			\hline
			bf&$E_0=\sqrt{1-p}I,E_1=\sqrt{p}\sigma_1$\\
			\hline
			pf&$E_0=\sqrt{1-p}I,E_1=\sqrt{p}\sigma_3$\\
			\hline 
			bpf&$E_0=\sqrt{1-p}I,E_1=\sqrt{p}\sigma_2$\\
			\hline 
			dep&$E_0=\sqrt{1-p}I,E_1=\sqrt{\frac{p}{3}}\sigma_1$\\
			&$E_2=\sqrt{\frac{p}{3}}\sigma_2,E_3=\sqrt{\frac{p}{3}}\sigma_3$\\
			\hline
			adc&$E_0=\left(\begin{array}{cc}
				1 & 0\\
				0 & \sqrt{1-p}
			\end{array}\right)$,
			$E_1=\left(\begin{array}{cc}
				0 & \sqrt{p}\\
				0 & 0
			\end{array}\right)$\\
			\hline
			dp&$E_0=\left(\begin{array}{cc}
				\sqrt{1-p} & 0\\
				0 & \sqrt{1-p}
			\end{array}\right)$,
			$E_1=\left(\begin{array}{cc}
				\sqrt{p} & 0\\
				0 & 0
			\end{array}\right)$,\\
			&$E_2=\left(\begin{array}{cc}
				0 & 0\\
				0 & \sqrt{p}
			\end{array}\right)$\\
			\hline 
		\end{tabular}
		
		\vspace{4pt}
		{\footnotesize
			\begin{minipage}{\textwidth}
				\textbf{Table I:} Kraus operators for the quantum channels: bit flip channel (bf), phase flip channel (pf), bit-phase flip channel (bpf), depolarizing channel (dep), amplitude damping channel (adc), and dephasing channel (dp), where $p$ are decoherence probabilities, $0 < p < 1$.
			\end{minipage}
		}
	\end{table*}
	Let $\mathbi{C}(\rho_{bf}^{'},H_{ABC})$, $\mathbi{C}(\rho_{pf}^{'},H_{ABC})$, $\mathbi{C}(\rho_{bpf}^{'},H_{ABC})$, $\mathbi{C}(\rho_{dep}^{'},H_{ABC})$, and $\mathbi{C}(\rho_{dp}^{'},H_{ABC})$ be quantum battery capacity of $\rho^{'}$ under bit flip channel, phase flip channel, bit-phase flip channel, depolarizing channel and dephasing channel, respectively.The detailed calculation process can be found in the Appendix B. And expression for battery capacity of $\mathbi{C}(\rho_{pf}^{'},H_{ABC})$, $\mathbi{C}(\rho_{bpf}^{'},H_{ABC})$ and $\mathbi{C}(\rho_{dep}^{'},H_{ABC})$ are listed in the Appendix B. From (\ref{e3}), we have
	\begin{equation*}
		\begin{split}
			\mathbi{C}(\rho_{bf}^{'},H_{ABC})&=2(1-3p+3p^2)(\epsilon^A+\epsilon^B+\epsilon^C)\\
			&+2(p-p^2)(3\epsilon^A-\epsilon^B-\epsilon^C)\\	
			\mathbi{C}(\rho_{dp}^{'},H_{ABC})&=(2-3p+3p^2-p^3)(\epsilon^A+\epsilon^B+\epsilon^C)\\
			&+(3p-3p^2+p^3)(\epsilon^A+\epsilon^B-\epsilon^C)\\
		\end{split}
	\end{equation*}
	
	Under the amplitude damping channel, the input state, i.e GHZ state, is transformed into the output state on the first subsystem, $\rho_{adc}^{'}=(E_0 \otimes I \otimes I)\rho(E_{0}^{\dagger} \otimes I \otimes I)+(E_1 \otimes I \otimes I)\rho(E_{1}^{\dagger} \otimes I \otimes I)$.
	$$ 
	\rho_{adc}^{'}=\frac{1}{2}\left(\begin{array}{cccccccc}
		1 & 0 & 0 & 0 & 0 & 0 & 0 & \sqrt{1-p}\\
		0 & 0 & 0 & 0 & 0 & 0 & 0 & 0\\
		0 & 0 & 0 & 0 & 0 & 0 & 0 & 0\\
		0 & 0 & 0 & p & 0 & 0 & 0 & 0\\
		0 & 0 & 0 & 0 & 0 & 0 & 0 & 0\\
		0 & 0 & 0 & 0 & 0 & 0 & 0 & 0\\
		0 & 0 & 0 & 0 & 0 & 0 & 0 & 0\\
		\sqrt{1-p} & 0 & 0 & 0 & 0 & 0 & 0 & 1-p
	\end{array}\right).
	$$\\
	So the eigenvalues of $\rho_{adc}^{'}$ are $\lambda_i=0$, (i=0,1,2,3,4,5), $\lambda_6=\frac{p}{2}$, $\lambda_7=1-\frac{p}{2}$. Therefore, the order of eigenvalues is $\lambda_0=\lambda_1=...=\lambda_5\leqslant \lambda_6\leqslant\lambda_7$. Set $\epsilon^A=0.5$, $\epsilon^B=0.3$, and $\epsilon^C=0.1$. Using (\ref{e1}),the detailed process can be found in Appendix B6, we have 

	\begin{equation*}
			\mathbi{C}(\rho_{adc}^{'},H_{ABC})=1.8-0.2p.
	\end{equation*}
	\\
	
	In Figure 1, We find that under the amplitude damping channel, as p increases, the quantum battery capacity monotonically decreases for initial state $\rho_{ABC}$; under the dephasing channel, the quantum battery capacity gradually decreases and approaches a constant with the increase of p, which is known as the frozen capacity phenomenon;  under the bit-phase flip channel, the quantum battery capacity monotonically decreases, when p = 0.5, a brief sudden death of capacity occurs,and then the quantum battery capacity monotonically increase with the increase of p. Similarly, under the  depolarizing channel, there is a phenomenon of a brief sudden capacity death when p = 0.75, followed by an increase in battery capacity. We observe significant differences in the evolution of battery capacity across various Markovian channels, indicating that the physical mechanism of noise is critical for the performance of quantum batteries. Compared to the depolarizing channel, the capacity of the quantum battery under the bit-phase flip channel decreases more rapidly, experiences the phenomenon of sudden capacity death at an earlier stage, and exhibits a stronger recovery ability. In comparison to other noise channels, the quantum battery capacity under the amplitude damping channel decreases monotonically with increasing noise intensity. This indicates that the storage capacity of the battery exhibits greater stability.
	\begin{figure}[htbp]
		\centering
		\includegraphics[width=0.5\textwidth]{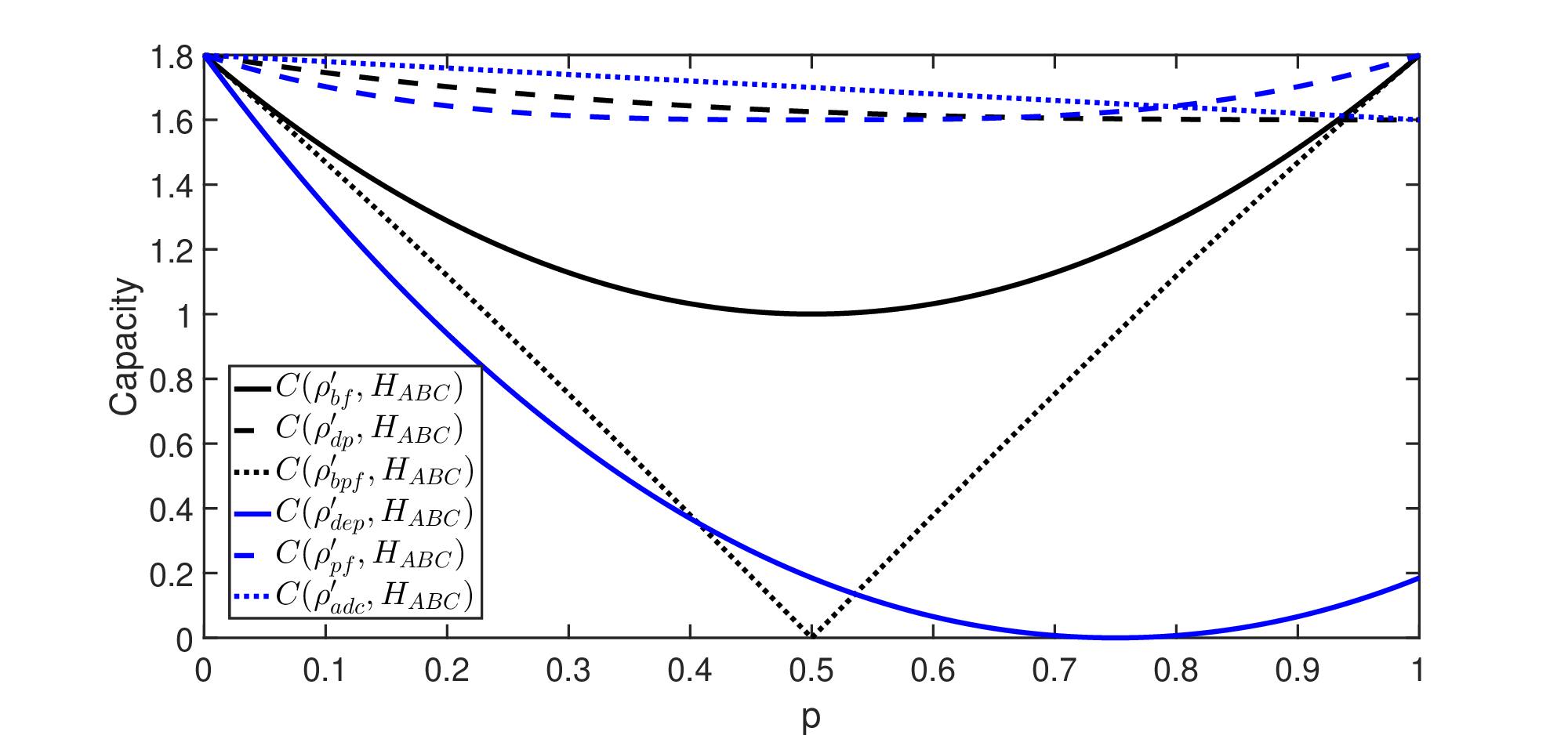}
		\vspace{-2em} \caption{ Quantum battery capacity evolution for GHZ state with $\epsilon^A=0.5, \epsilon^B=0.3, \epsilon^C=0.1$ under bit flip channel ($\mathbi{C}(\rho_{bf}^{'},H_{ABC})$), phase flip channel ($\mathbi{C}(\rho_{pf}^{'},H_{ABC})$), bit-phase flip channel ($\mathbi{C}(\rho_{bpf}^{'},H_{ABC})$), depolarizing channel ($\mathbi{C}(\rho_{dep}^{'},H_{ABC})$), depasing channel ($\mathbi{C}(\rho_{dp}^{'},H_{ABC})$) and amplitude damping channel ($\mathbi{C}(\rho_{adc}^{'},H_{ABC})$) as a function of p.} \label{Fig.1}
	\end{figure}
	\subsection{B. Quantum battery capacity evolution under n times Markovian channels on the first system}
	Next, we consider the quantum battery capacity variation of GHZ state under n times Markovian channels. We find that phase flip channel and dephasing channel preserve the form of GHZ state.

	Let $\mathbi{C}(\rho_{pf}^{(n)},H_{ABC})$, $\mathbi{C}(\rho_{dp}^{(n)},H_{ABC})$ and $\mathbi{C}(\rho_{adc}^{(n)},H_{ABC})$ be quantum battery capacity under n times phase flip channel, n times dephasing channel and n times amplitude damping channel, respectively. We present only the battery capacity expressions for the n times phase flip channel and dephasing channel, while omitting a comprehensive discussion of their derivation. The detailed derivation process can be found in Appendix C. From (\ref{e3}), we have
	\begin{equation*}
		\begin{split}
			&\mathbi{C}(\rho_{pf}^{(n)},H_{ABC})\\
			&=(1+\sqrt{(1-2p)^{6n}})\times 0.9+(1-\sqrt{(1-2p)^{6n}})\times 0.7
		\end{split}
	\end{equation*}
	\begin{equation*}
		\begin{split}
			&\mathbi{C}(\rho_{dp}^{(n)},H_{ABC})\\
			&=(1+\sqrt{(1-p)^{6n}})\times 0.9+(1-\sqrt{(1-p)^{6n}})\times 0.7.
		\end{split}
	\end{equation*}
	
	\begin{figure*}[htbp]
		\centering
		\begin{minipage}[b]{0.48\linewidth}
			\centering
			\includegraphics[width=\linewidth]{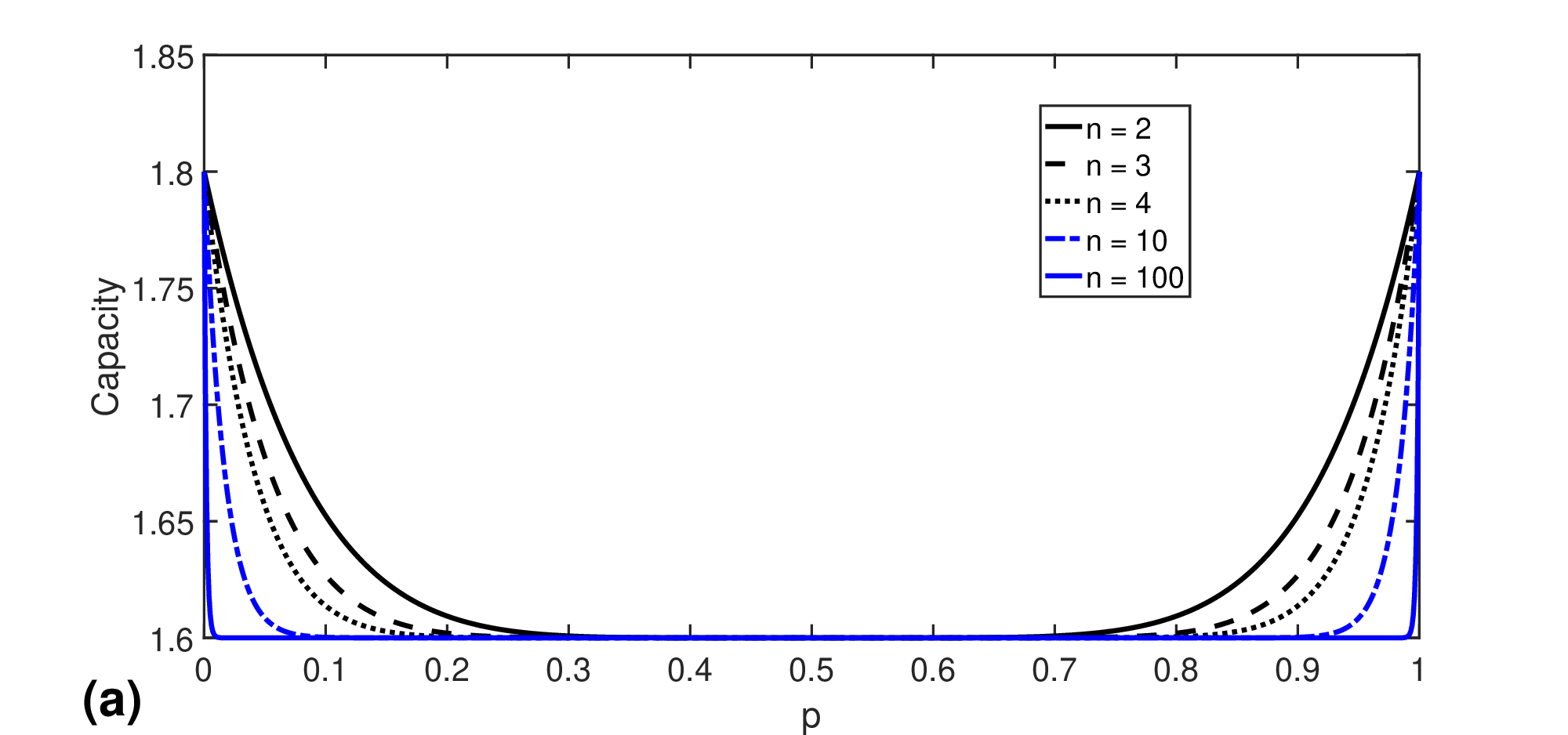}
			\label{Fig2a:left_mini}
		\end{minipage}
		\hfill
		\begin{minipage}[b]{0.48\linewidth}
			\centering
			\includegraphics[width=\linewidth]{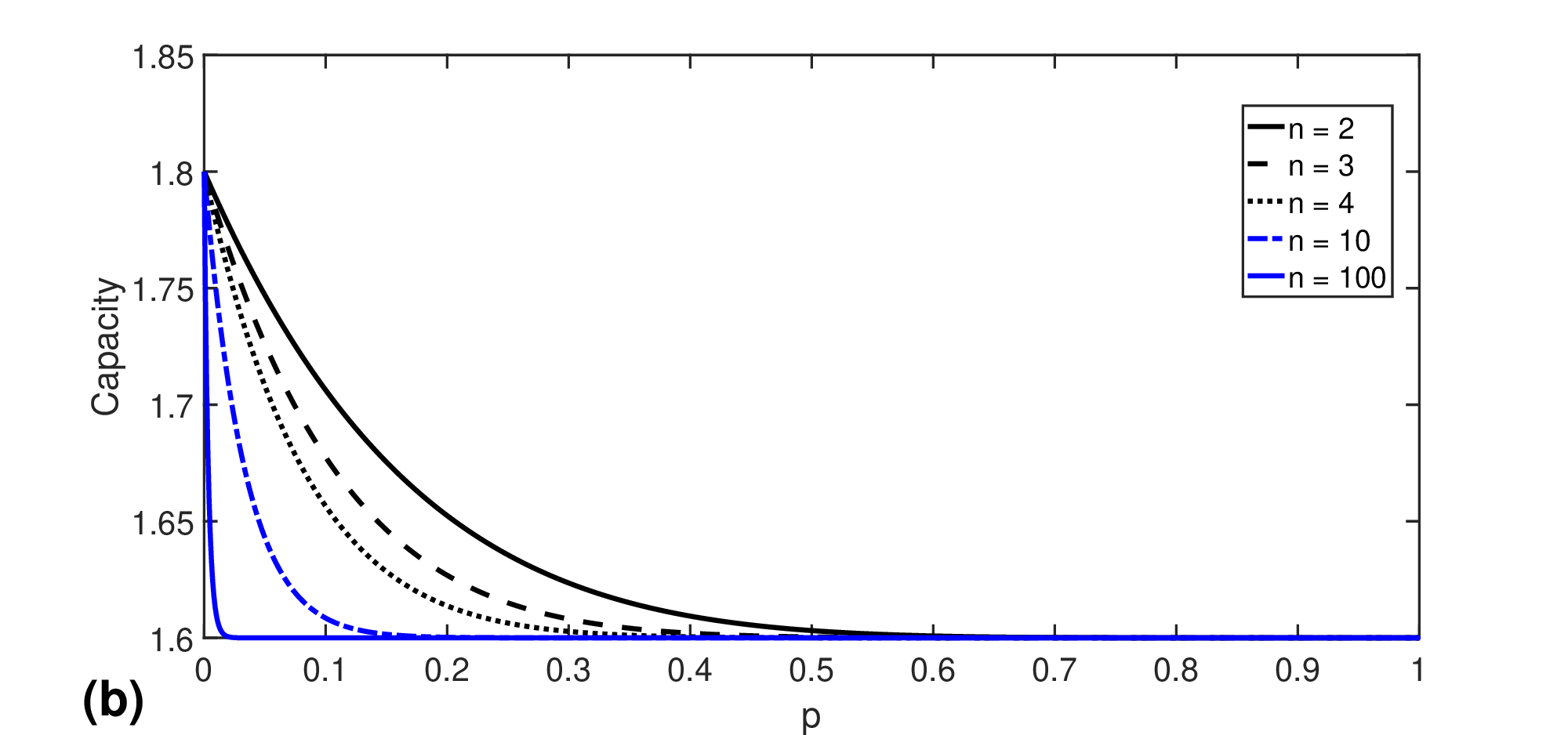}
			\label{Fig2b:right_mini}
		\end{minipage}
		\caption{(a) Quantum battery capacity evolution for GHZ state with $\epsilon^A=0.5$, $\epsilon^B=0.3$ and $\epsilon^c=0.1$ under the phase flip channel $n$ times as a function of $p$. (b) Quantum battery capacity evolution for GHZ state with $\epsilon^A=0.5$, $\epsilon^B=0.3$ and $\epsilon^c=0.1$ under the dephasing channel $n$ times as a function of $p$.}
		\label{Fig2:minipage_example}
	\end{figure*}
	Figure 2 (a) shows the change in quantum battery capacity under n times phase flip channel, and it can be observed that as p increases, the battery capacity undergoes a process of decay and recovery, during which non-Markovian behavior is exhibited. Figure 2 (b) shows the variation of quantum battery capacity under n times dephasing channels. We can observe the occurrence of frozen capacity phenomenon, and the larger n, the earlier the frozen capacity phenomenon occurs. The phenomenon of frozen capacity indicates that even in the presence of continuous noise or multiple noise interference, battery capacity does not continuously decay to zero. Instead, it stabilizes at a constant non-zero value, allowing quantum batteries to maintain a certain level of energy storage capacity.

	In addition, after going through the amplitude damping channel n times on the first subsystem, the capacity of the output state $\rho_{adc}^{(n)}$ in Appendix C3 is the 
	\begin{equation*}
			\mathbi{C}(\rho_{adc}^{(n)},H_{ABC})
			=0.2\sqrt{(1-p)^{2n}}+1.6.		
	\end{equation*}
	\begin{figure}[htbp]
		\centering
		\includegraphics[width=0.5\textwidth]{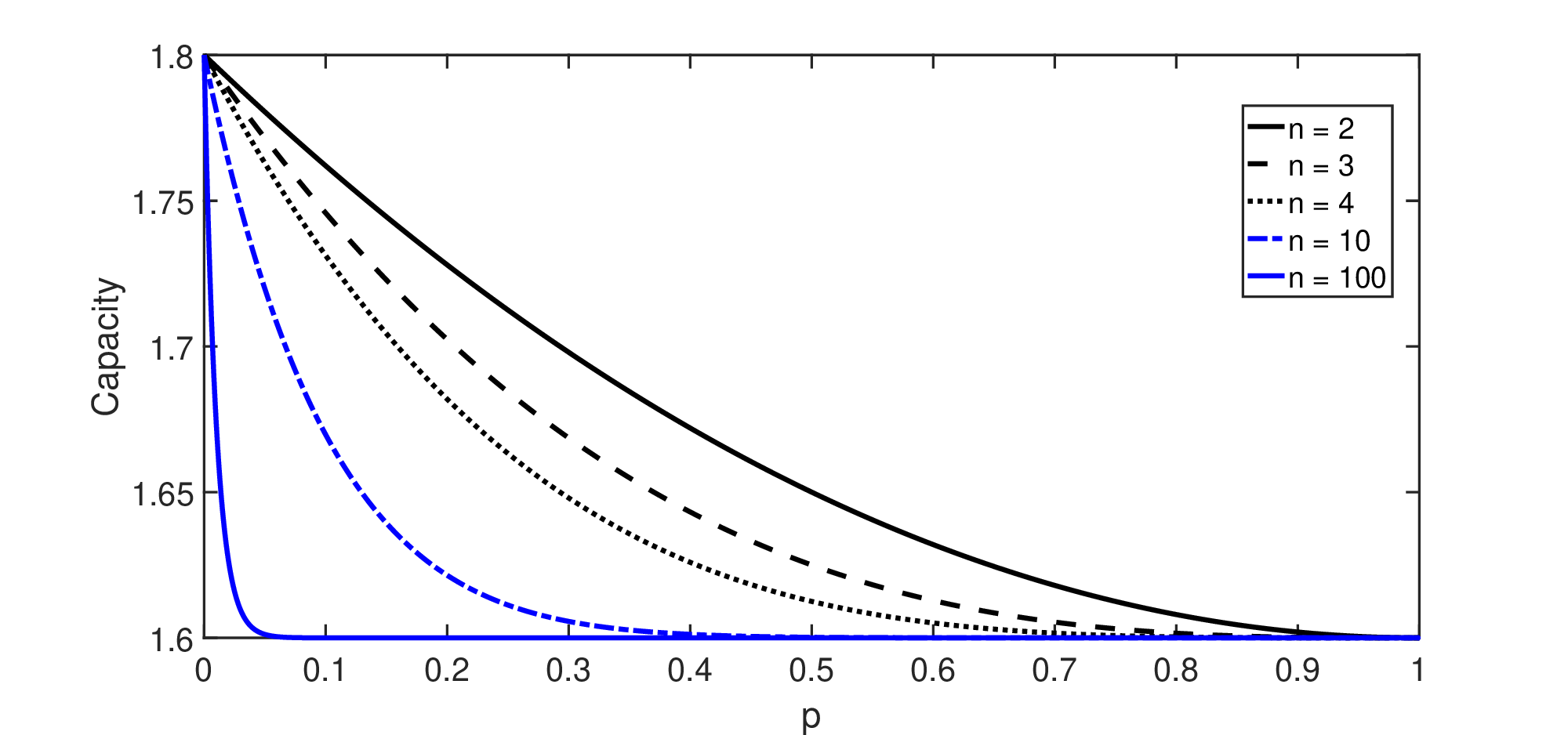}
		\vspace{-2em} \caption{ Quantum battery capacity evolution for GHZ state with $\epsilon^A=0.5, \epsilon^B=0.3, \epsilon^C=0.1$ under the amplitude damping channel $n$ times on the first subsystem as a function of $p$.}\label{Fig.3}
	\end{figure}

    Figure 3 shows the change in quantum battery capacity under n times the amplitude damping channel on the first subsystem. Similar to the dephasing channel, a capacity freezing phenomenon also occurs in the amplitude damping channel, and this phenomenon appears earlier as n increases. A comparison between the two reveals that in the amplitude damping channel, the rate of capacity decay of the quantum battery is slower, and the capacity freezing phenomenon occurs later. This indicates that the storage capacity in the amplitude damping channel is superior to that in the dephasing channel.
    
	\subsection{C. Quantum battery capacity evolution of GHZ state under tri-side Markovian channels of the same type.}
	In this part, we mainly investigate the variation of quantum battery capacity for GHZ state undergoing three independent local Markovian channels of the same type, see the Kraus operators in Table II.
	\begin{table*}[htbp] 
		\centering
		\begin{tabular}{|c|c|}
			\hline
			channel& Kraus operators \\
			\hline
			bf-bf-bf&$E_{0}^{A}=\sqrt{1-p}I^A\otimes I^B\otimes I^C,E_{1}^{A}=\sqrt{p}\sigma_{1}^{A}\otimes I^B\otimes I^C$\\			
			&$E_{0}^{B}=I^A\otimes \sqrt{1-q}I^B\otimes I^C,E_{1}^{B}=I^A\otimes \sqrt{q}\sigma_{1}^{B}\otimes I^C$\\
			&$E_{0}^{C}=I^A\otimes I^B\otimes \sqrt{1-\gamma}I^C,E_{1}^{C}=I^A\otimes I^B\otimes \sqrt{\gamma}\sigma_{1}^{C}$\\
			\hline
			pf-pf-pf&$E_{0}^{A}=\sqrt{1-p}I^A\otimes I^B\otimes I^C,E_{1}^{A}=\sqrt{p}\sigma_{3}^{A}\otimes I^B\otimes I^C$\\
			&$E_{0}^{B}=I^A\otimes \sqrt{1-q}I^B\otimes I^C,E_{1}^{B}=I^A\otimes \sqrt{q}\sigma_{3}^{B}\otimes I^C$\\
			&$E_{0}^{C}=I^A\otimes I^B\otimes \sqrt{1-\gamma}I^C,E_{1}^{C}=I^A\otimes I^B\otimes \sqrt{\gamma}\sigma_{3}^{C}$\\
			\hline 
			bpf-bpf-bpf&$E_{0}^{A}=\sqrt{1-p}I^A\otimes I^B\otimes I^C,E_{1}^{A}=\sqrt{p}\sigma_{2}^{A}\otimes I^B\otimes I^C$\\
			&$E_{0}^{B}=I^A\otimes \sqrt{1-q}I^B\otimes I^C,E_{1}^{B}=I^A\otimes \sqrt{q}\sigma_{2}^{B}\otimes I^C$\\
			&$E_{0}^{C}=I^A\otimes I^B\otimes \sqrt{1-\gamma}I^C,E_{1}^{C}=I^A\otimes I^B\otimes \sqrt{\gamma}\sigma_{2}^{C}$\\
			\hline 
		\end{tabular}
		
		\vspace{4pt}
		{\footnotesize
			\begin{minipage}{\textwidth}
				\textbf{Table II:} Kraus operators for three independent local Markovian channels: three independent local bit flip channel (bf-bf-bf), three independent local phase flip channel (pf-pf-pf), three independent local bit-phase flip channel (bpf-bpf-bpf) where $p,q,\gamma$ are decoherence probabilities, $0 < p,q,\gamma < 1.$
			\end{minipage}
		}
	\end{table*}
	
	According to Eq.(\ref{e4}) the initial state $\rho_{ABC}$ can evolve to another state $\rho^{'}=\varepsilon(\rho)$ under the Markovian channels. From (\ref{e3}), we have the dynamical behaviours of the quantum battery capacity for GHZ state under the tri-side same type phase flip channel, see Fig. 4. The quantum battery capacity of GHZ state under the phase flip channel is shown in the Appendix D1. Assuming that $\gamma$ is equal to 0.25, 0.5, 0.75, and 1, respectively. When the three subsystems experience phase flip channels of the same type but varying intensities, we can observe that the capacity evolution behaviour of the GHZ state exhibits significant differences, yet it generally maintains a high level. Notably, when $\gamma = 0.5$ , the capacity remains fixed regardless of the values of p and q. In comparison to the other three scenarios, the performance of the quantum battery capacity is more stable.
	   
\begin{figure*}[htbp]
	\centering
	\begin{minipage}[b]{0.48\linewidth}
		\centering
		\includegraphics[width=\linewidth]{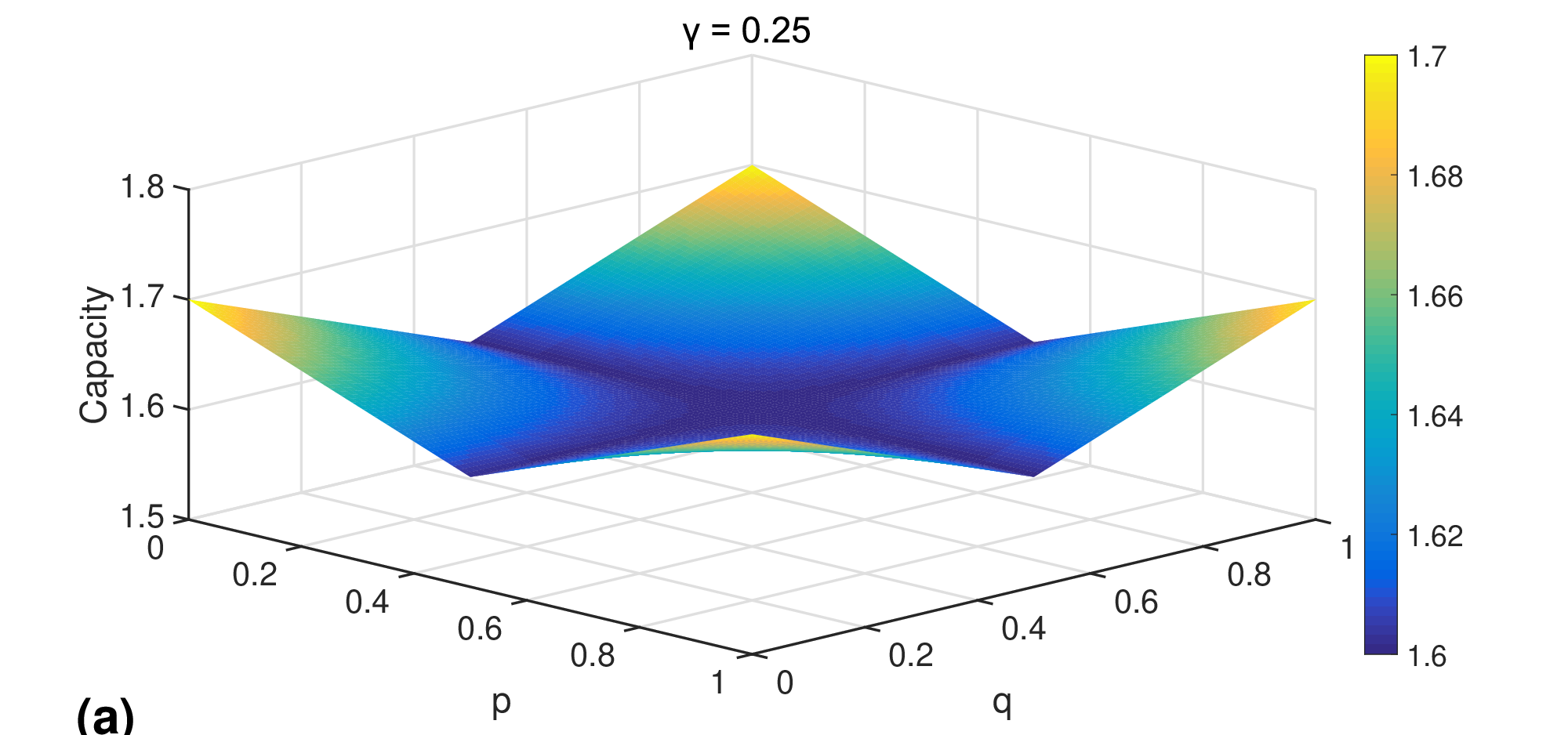}
		\label{Fig.4a}
	\end{minipage}
	\hfill
	\begin{minipage}[b]{0.48\linewidth}
		\centering
		\includegraphics[width=\linewidth]{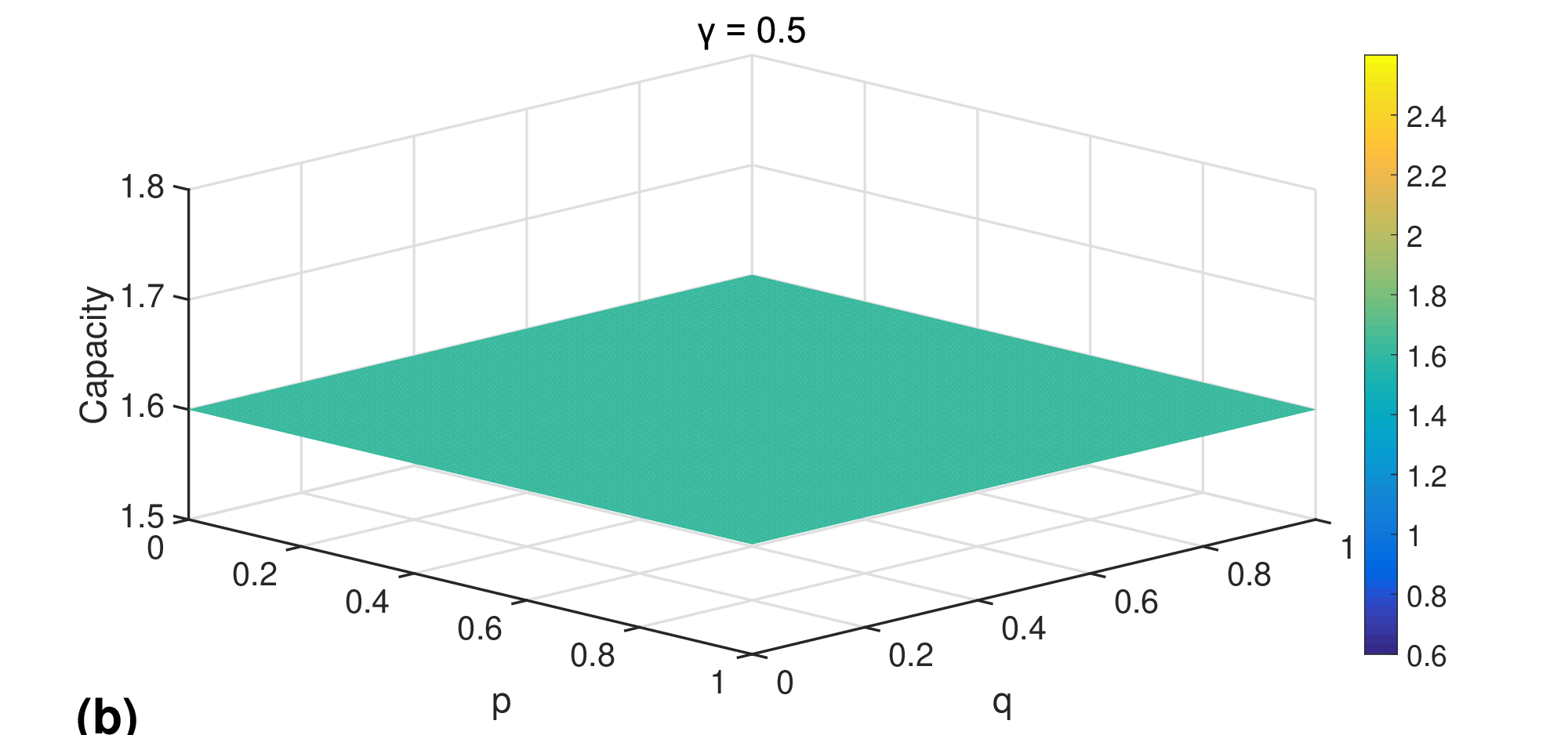}
		\label{Fig.4b}
	\end{minipage}
	\vspace{0.5em}
	\begin{minipage}[b]{0.48\linewidth}
		\centering
		\includegraphics[width=\linewidth]{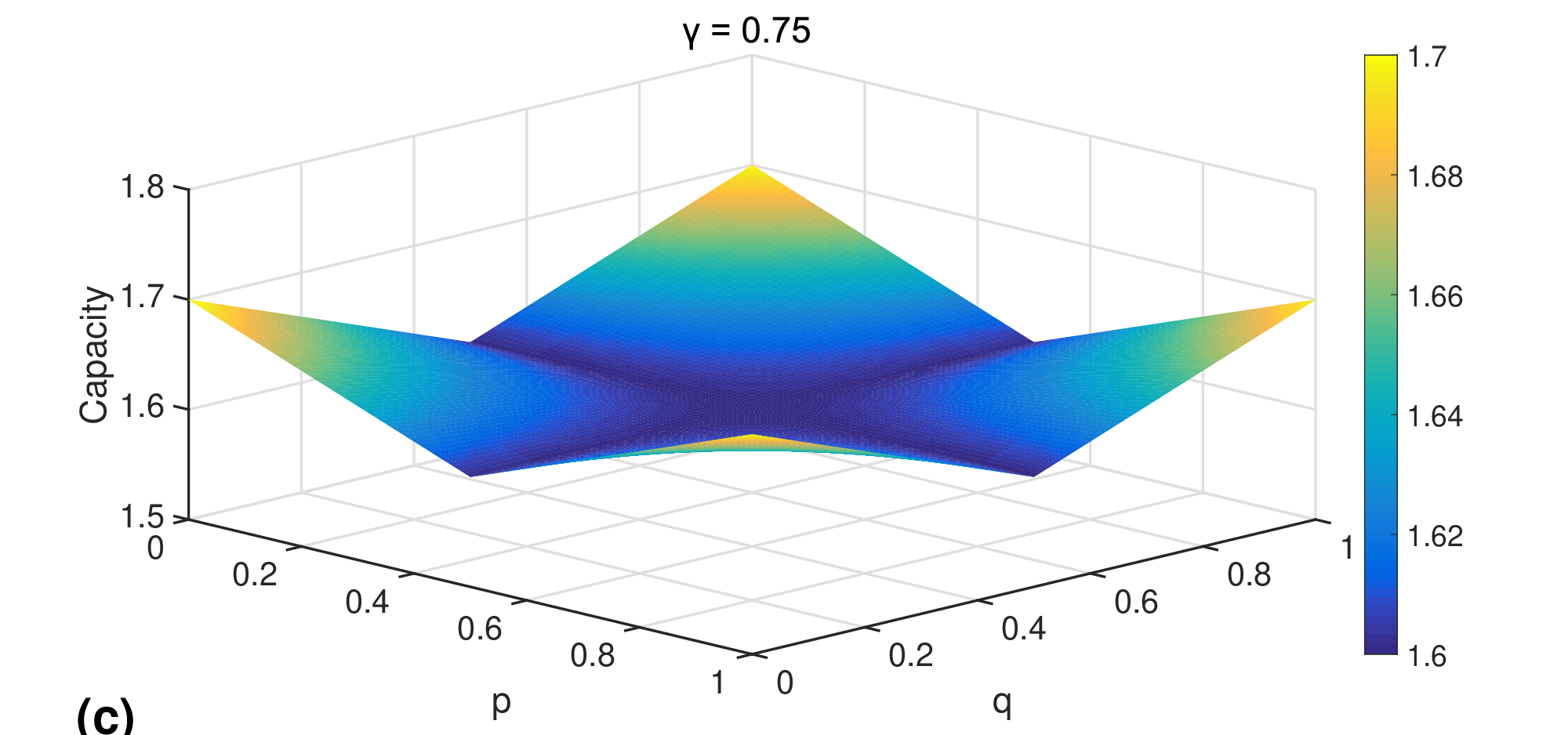}
		\label{Fig.4c}
	\end{minipage}
	\hfill
	\begin{minipage}[b]{0.48\linewidth}
		\centering
		\includegraphics[width=\linewidth]{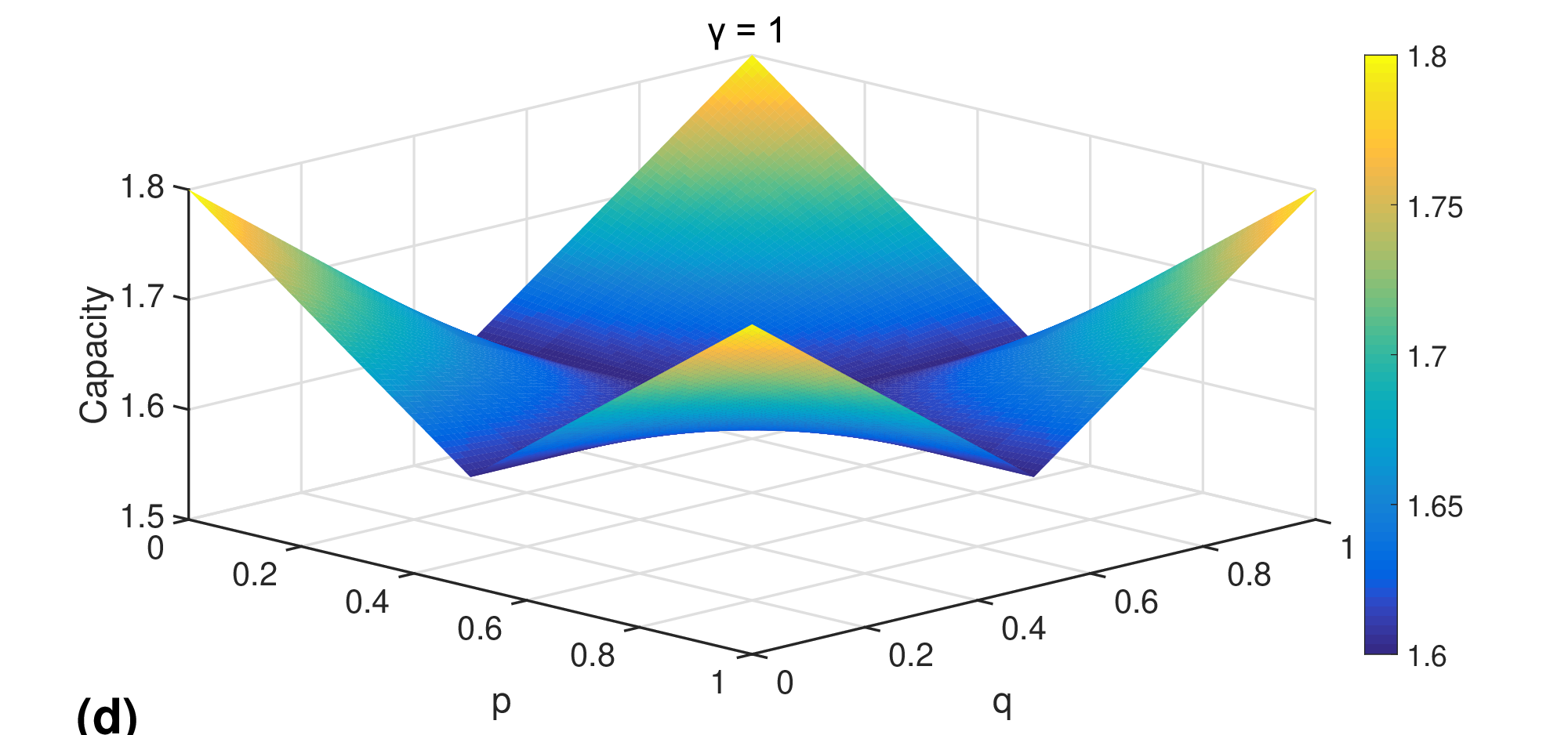}
		\label{Fig.4d}
	\end{minipage}
	
	\caption{Quantum battery capacity evolution for GHZ state with $\epsilon^A=0.5$, $\epsilon^B=0.3$ and $\epsilon^C=0.1$ under tri-side same type phase flip channel. (a) $\gamma=0.25$ (b) $\gamma=0.5$ (c) $\gamma=0.75$ (d) $\gamma=1$
	}
	\label{Fig.4}
\end{figure*}
	
	Furthermore, if all subsystems go through the same type channel n times, the initial state $\rho_{ABC}$  also evolves to another state $\rho^{'}=\varepsilon(\rho)$ under the Markovian channels. By (\ref{e3}), we can observe the dynamical behaviours of the quantum battery capacity for GHZ state under the tri-side same type phase flip channel n times, see Fig. 5. The quantum battery capacity of GHZ state under the phase flip channel is shown in the Appendix D2. Assuming that $\gamma$ is equal to 0.25, 0.5, 0.75, and 1, respectively. We can observe significant differences in the capacity evolution behaviour of GHZ states under n times tri-side phase-flip channels. However, even after passing through multiple noisy channels, the capacity remains at a relatively high level. Notably, when $\gamma = 0.5$, the capacity maintains a constant value regardless of the values of p and q. Compared to the other three scenarios, it can be concluded that the storage capacity of quantum batteries is not affected by the intensity of noise, demonstrating a strong stability.
	
	\begin{figure*}[htbp]
		\centering
		\begin{minipage}[b]{0.48\linewidth}
			\centering
			\includegraphics[width=\linewidth]{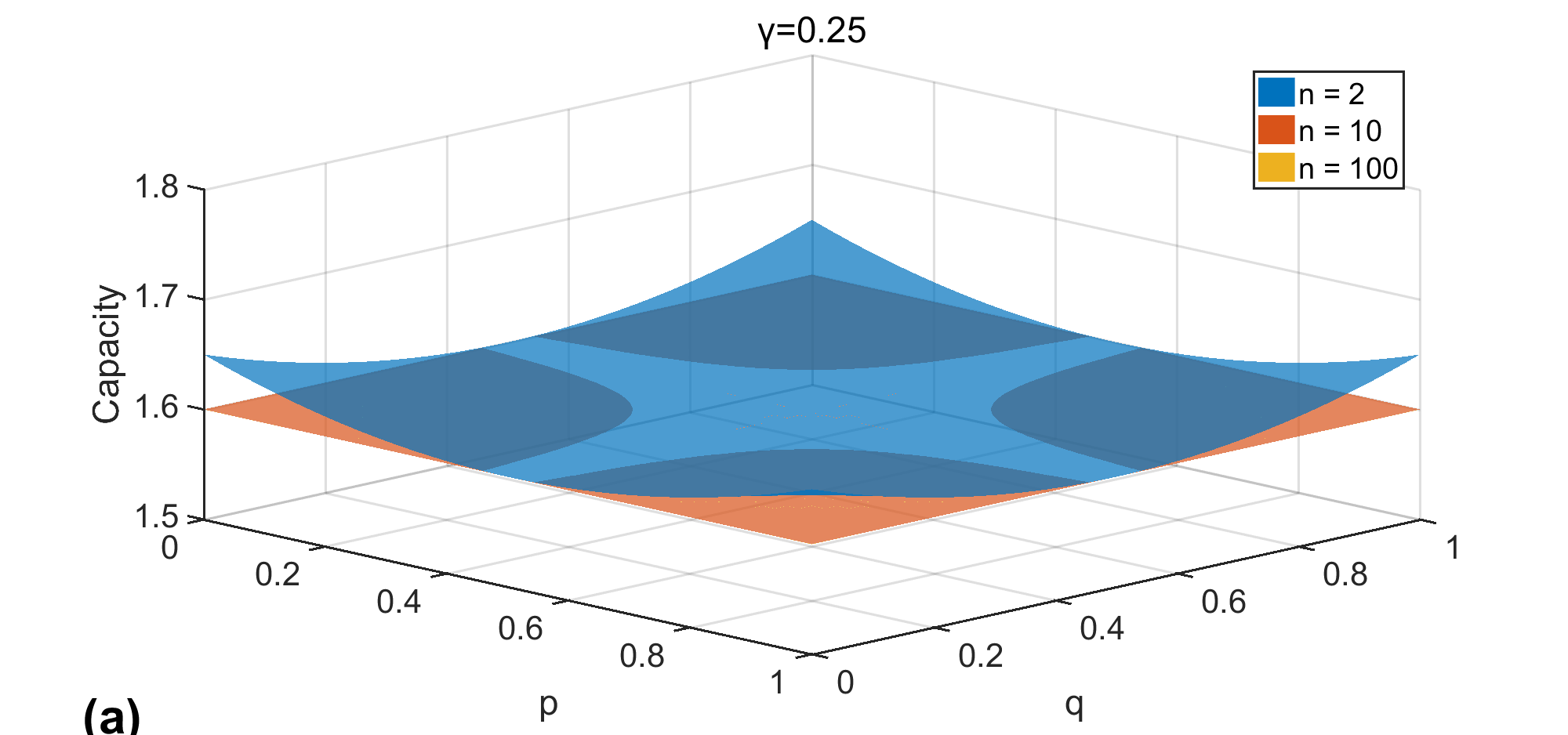}
			\label{Fig.5a}
		\end{minipage}
		\hfill
		\begin{minipage}[b]{0.48\linewidth}
			\centering
			\includegraphics[width=\linewidth]{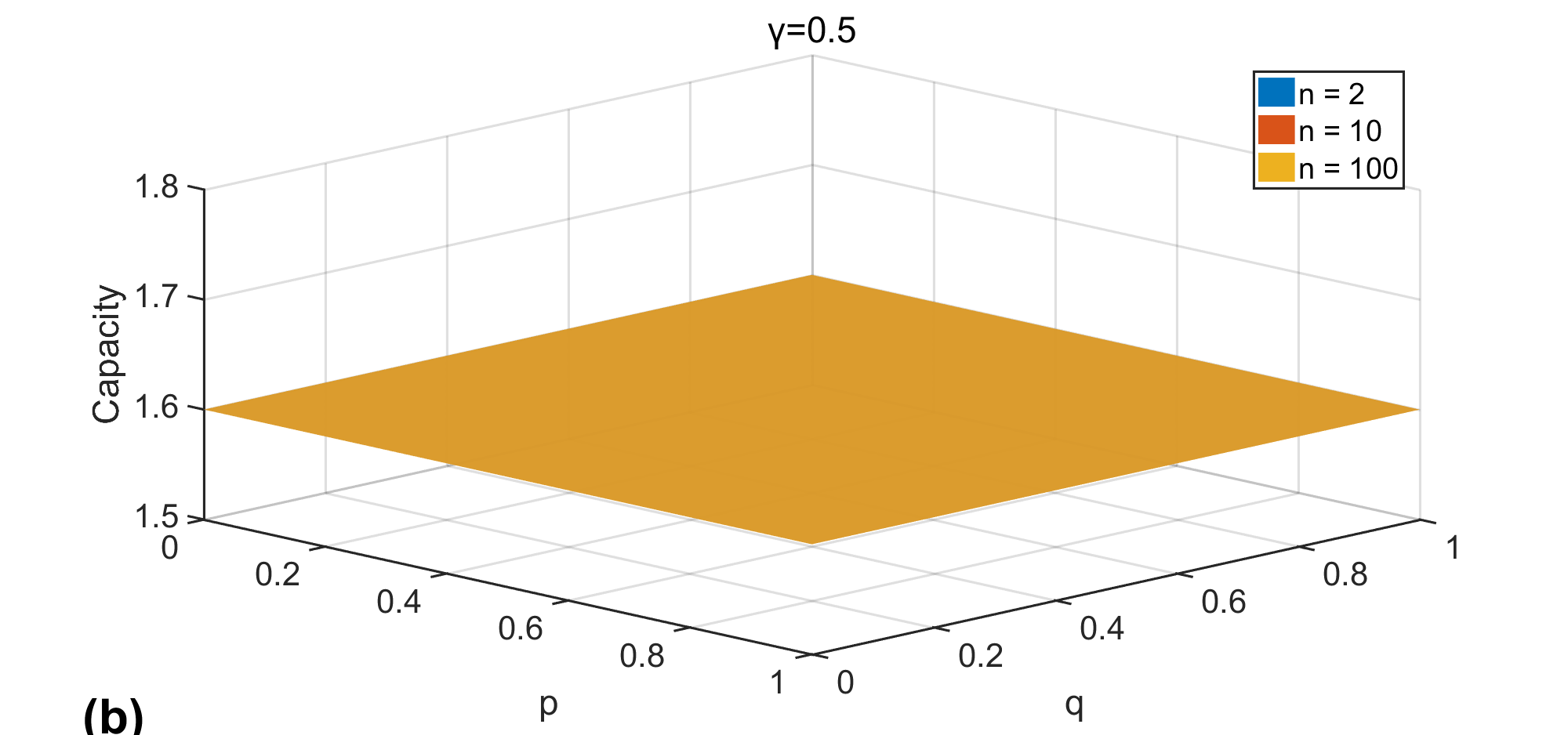}
			\label{Fig.5b}
		\end{minipage}
		\vspace{0.5em}
		\begin{minipage}[b]{0.48\linewidth}
			\centering
			\includegraphics[width=\linewidth]{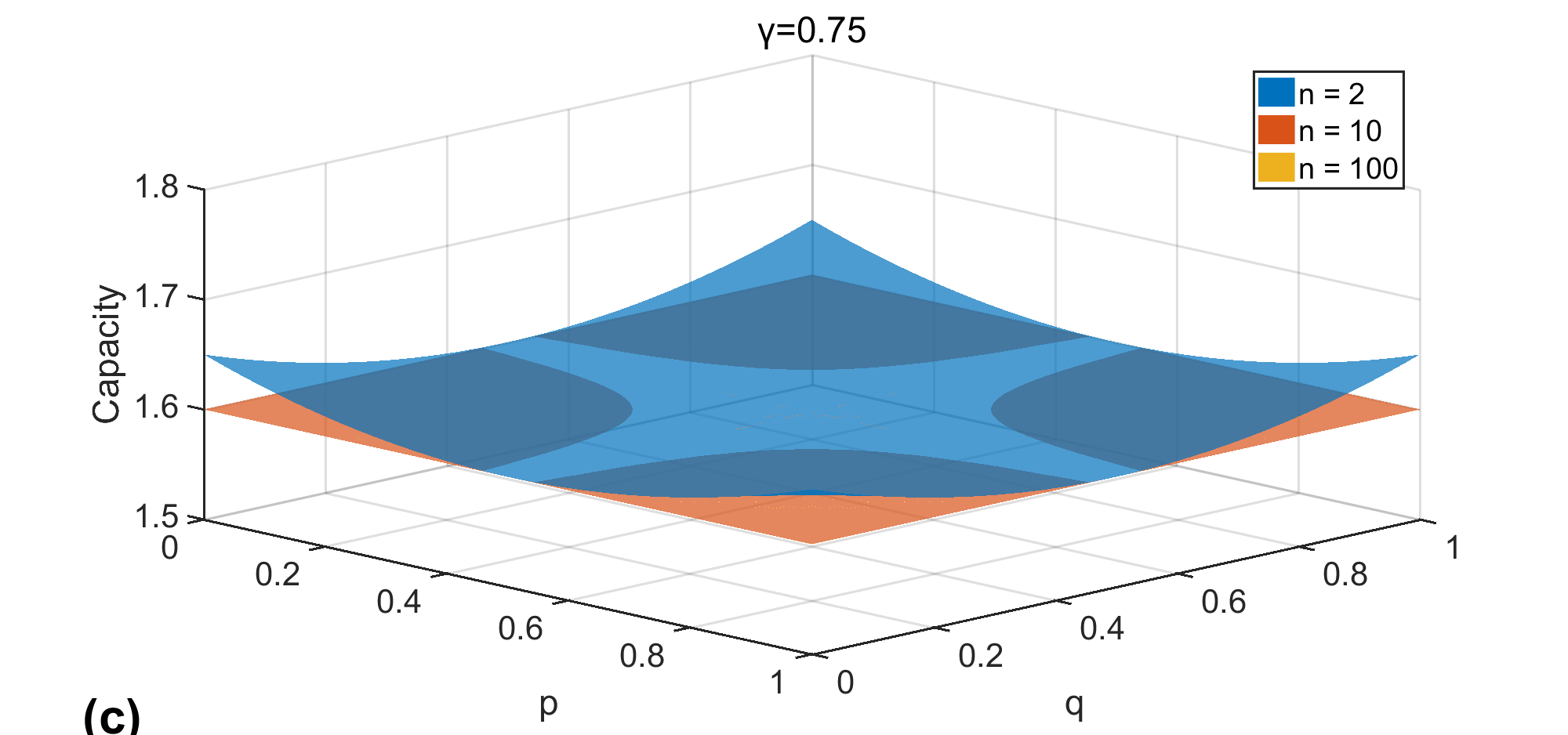}
			\label{Fig.5c}
		\end{minipage}
		\hfill
		\begin{minipage}[b]{0.48\linewidth}
			\centering
			\includegraphics[width=\linewidth]{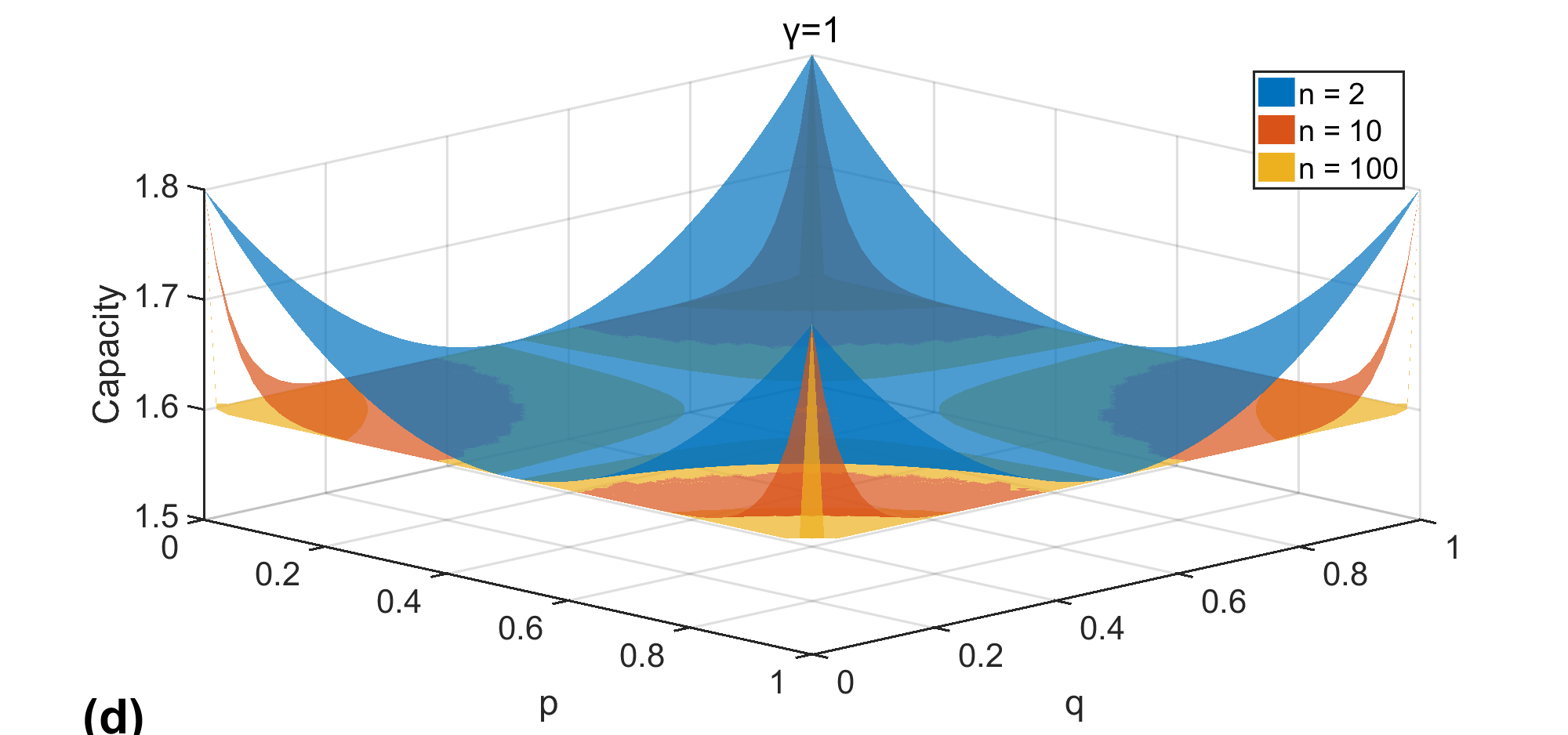}
			\label{Fig.5d}
		\end{minipage}
		
		\caption{ Quantum battery capacity evolution for GHZ state with $\epsilon^A=0.5$, $\epsilon^B=0.3$ and $\epsilon^C=0.1$ under tri-side same type phase flip channel $n$ times. (a) $\gamma=0.25$  (b) $\gamma=0.5$ (c) $\gamma=0.75$ (d) $\gamma=1$
		}
		\label{Fig.5}
	\end{figure*}
	
	\section{III. Quantum battery capacity of GHZ-like states in tripartite system under Markovian channels}
	We consider the three-qubit GHZ-like states,
	\begin{equation}
		\begin{split}\label{e5}
			\rho=|GHZ\rangle_{like}\langle GHZ|
		\end{split}
	\end{equation}
	where $|GHZ\rangle_{like}=a|000\rangle+\sqrt{1-a^2}|111\rangle$, a is a real constant, $a\in[0,1]$ and $a^2+(\sqrt{1-a^2})^2=1$. The eigenvalues of $\rho$ are $\lambda_i=0, (i=0,...,6)$ and $\lambda_7=1$. Therefore, from (\ref{e1}), we have
	$$
	\mathbi{C}(\rho,H_{ABC})=2(\epsilon^A+\epsilon^B+\epsilon^C).
	$$
	To simplify the expression of eigenvalues and battery capacity, we assume $c_1=a^2$, $c_2=a\sqrt{1-a^2}$ and $c_3=1-a^2$, the (\ref{e5}) can be rewritten as
	$$
	\rho=c_1|000\rangle\langle000|+c_2|000\rangle\langle111|
	+c_2|111\rangle\langle000|+c_3|111\rangle\langle111|.
	$$
	The eigenvalues of $\rho$ are $\lambda_i=0, (i=0,...,5)$, $\lambda_6=\frac{c_1+c_3-\sqrt{(c_1-c_3)^2+4c_{2}^{2}}}{2}$ and $\lambda_7=\frac{c_1+c_3+\sqrt{(c_1-c_3)^2+4c_{2}^{2}}}{2}$. Therefore, from (\ref{e1}), the quantum battery capacity of $\rho$ is
	\begin{equation}\label{e6}
		\begin{split}
			\mathbi{C}(\rho,H_{ABC})&=(c_1+c_3+\sqrt{(c_1-c_3)^2+4c_{2}^{2}})(\epsilon^A+\epsilon^B+\epsilon^C)\\
			&=(c_1+c_3-\sqrt{(c_1-c_3)^2+4c_{2}^{2}})(\epsilon^A+\epsilon^B-\epsilon^C).
		\end{split}
	\end{equation}
	\subsection{A. Quantum battery capacity evolution of GHZ-like states under Markovian channels on the first subsystem}
	In this section, we also mainly study the evolution of quantum state $\rho$ under a trace-preserving quantum operation $\varepsilon(\rho)$ \cite{mail},
	$$ 
	\varepsilon(\rho)=\sum_{i,j,k}(E_i \otimes E_j \otimes E_k)\rho(E_i \otimes E_j \otimes E_k)^{\dagger},
	$$
	where $\{E_m\}$ is the set of Kraus operators and satisfies completeness relationship, i.e $\sum_{m}E_m^{\dagger}E_m=I$. Several kraus operators of typical channel are listed in Table I, where phase flip channel (pf) and dephasing channel (dp) preserve the form of the density operator $\rho$. In this situation one has
	\begin{equation}\label{e7}
		\begin{split}
			\rho^{'}&=\varepsilon(\rho)\\
			&=c_{1}^{'}|000\rangle\langle000|+c_{2}^{'}|000\rangle\langle111|\\
			&+c_{2}^{'}|111\rangle\langle000|+c_{3}^{'}|111\rangle\langle111|,
		\end{split}
	\end{equation}
	where $c_{1}^{'}, c_{2}^{'}, c_{3}^{'} \in[0,1]$ are given in Table III.
	\begin{table}[htbp]
		\centering
		\begin{tabular}{|c|c|c|c|}
			\hline
			channel  & $c_{1}^{'}$  & $c_{2}^{'}$  & $c_{3}^{'}$  \\ \hline
			pf  & $c_1$  & $c_2(1-2p)^3$  & $c_3$  \\ \hline
			dp  & $c_1$  & $c_2(1-p)^3$  & $c_3$  \\ \hline
		\end{tabular}\\
		\vspace{4pt}
		{\footnotesize
			\textbf{Table III:} Correlation coefficients of the quantum operations: phase flip channel(pf) and dephasing channel(dp).
		}
	\end{table}
	\\
	Let $\mathbi{C}(\rho_{pf}^{'},H_{ABC})$ and $\mathbi{C}(\rho_{pf}^{'},H_{ABC})$ be quantum battery capacity of $\rho^{'}$ under phase flip channel and dephasing channel, respectively. The detailed calculation process can be found in the Appendix E. By (\ref{e6}), we have 
	\begin{equation*}
		\begin{split}
			&\mathbi{C}(\rho_{pf}^{'},H_{ABC})\\
			&=(c_1+c_3+\sqrt{(c_1-c_3)^2+4c_{2}^{2}(1-2p)^6})(\epsilon^A+\epsilon^B+\epsilon^C)\\
			&=(c_1+c_3-\sqrt{(c_1-c_3)^2+4c_{2}^{2}(1-2p)^6})(\epsilon^A+\epsilon^B-\epsilon^C),\\
			&\mathbi{C}(\rho_{dp}^{'},H_{ABC})\\
			&=(c_1+c_3+\sqrt{(c_1-c_3)^2+4c_{2}^{2}(1-p)^6})(\epsilon^A+\epsilon^B+\epsilon^C)\\
			&=(c_1+c_3-\sqrt{(c_1-c_3)^2+4c_{2}^{2}(1-p)^6})(\epsilon^A+\epsilon^B-\epsilon^C).
		\end{split}
	\end{equation*}
	\\
	Under the amplitude damping channel on the first subsystem, the GHZ-like state is transformed into the output state,
	$$
	\rho_{adc}^{'}=\left(\begin{array}{cccccccc}
		c_1 & 0 & 0 & 0 & 0 & 0 & 0 & c_2(1-p)^{\frac{1}{2}}\\
		0 & 0 & 0 & 0 & 0 & 0 & 0 & 0\\
		0 & 0 & 0 & 0 & 0 & 0 & 0 & 0\\
		0 & 0 & 0 & c_3p & 0 & 0 & 0 & 0\\
		0 & 0 & 0 & 0 & 0 & 0 & 0 & 0\\
		0 & 0 & 0 & 0 & 0 & 0 & 0 & 0\\
		0 & 0 & 0 & 0 & 0 & 0 & 0 & 0\\
		c_2(1-p)^{\frac{1}{2}} & 0 & 0 & 0 & 0 & 0 & 0 & c_3(1-p)
	\end{array}\right).
	$$
	The eigenvalues of $\rho_{adc}^{'}$ are $\lambda_i=0, i=0,...,4$, $\lambda_5=c_3p$, $\lambda_6=\frac{c_1+c_3(1-p)-\sqrt{(c_1-c_3+c_3)^2+4c_{2}^{2}(1-p)}}{2}$ and $\lambda_7=\frac{c_1+c_3(1-p)+\sqrt{(c_1-c_3+c_3)^2+4c_{2}^{2}(1-p)}}{2}$. Now let's assume two scenarios for $c_1 = a^2$. The first scenario is when a = 0.25, from (\ref{e6}), we have
	\begin{equation*}
		\begin{split}
		&\mathbi{C}(\rho_{adc}^{'},H_{ABC})\\
		&=\begin{cases}
			2\lambda_7(\epsilon^A+\epsilon^B+\epsilon^C)+2\lambda_5(\epsilon^A+\epsilon^B-\epsilon^C),  &p\in[0,\frac{2}{3}]\\
			2\lambda_5(\epsilon^A+\epsilon^B+\epsilon^C)+2\lambda_7(\epsilon^A+\epsilon^B-\epsilon^C),  &p\in(\frac{2}{3},1].
		\end{cases}
		\end{split}
	\end{equation*}
	 In the second scenario, when a = 0.75, according to (6), we have
	 \begin{equation*}
	 	\begin{split}
	 		\mathbi{C}(\rho_{adc}^{'},H_{ABC})
	 		=2\lambda_7(\epsilon^A+\epsilon^B+\epsilon^C)+2\lambda_5(\epsilon^A+\epsilon^B-\epsilon^C).
	 	\end{split}
	 \end{equation*}
	We find significant differences in the capacity evolution behaviour of GHZ-like states under three different Markovian channels, see figure 6. In particular, under the amplitude damping channel, when $a^2=0.25$, the battery capacity reaches a minimum value and then quickly rebounds; when $a^2=0.75$, the battery capacity monotonically decreases. It is reassuring that there is no sudden death of capacity. A comprehensive comparison of the three channels indicates that the capacity storage capability of the quantum battery is more stable under the dephasing channel.
	
	\begin{figure*}[htbp]
		\centering
		\begin{minipage}[b]{0.48\linewidth}
			\centering
			\includegraphics[width=\linewidth]{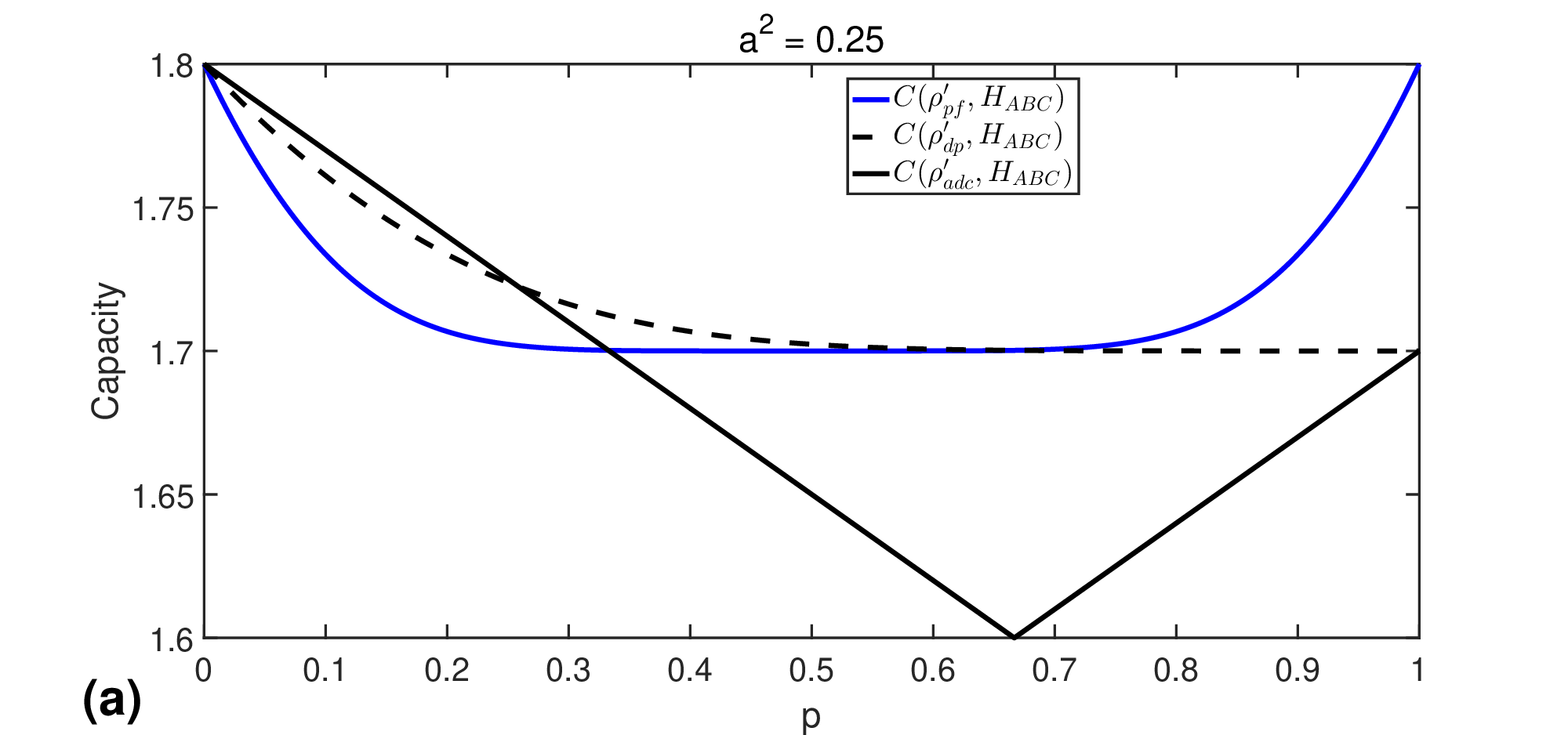}
			\label{Fig6a:left_mini}
		\end{minipage}
		\hfill
		\begin{minipage}[b]{0.48\linewidth}
			\centering
			\includegraphics[width=\linewidth]{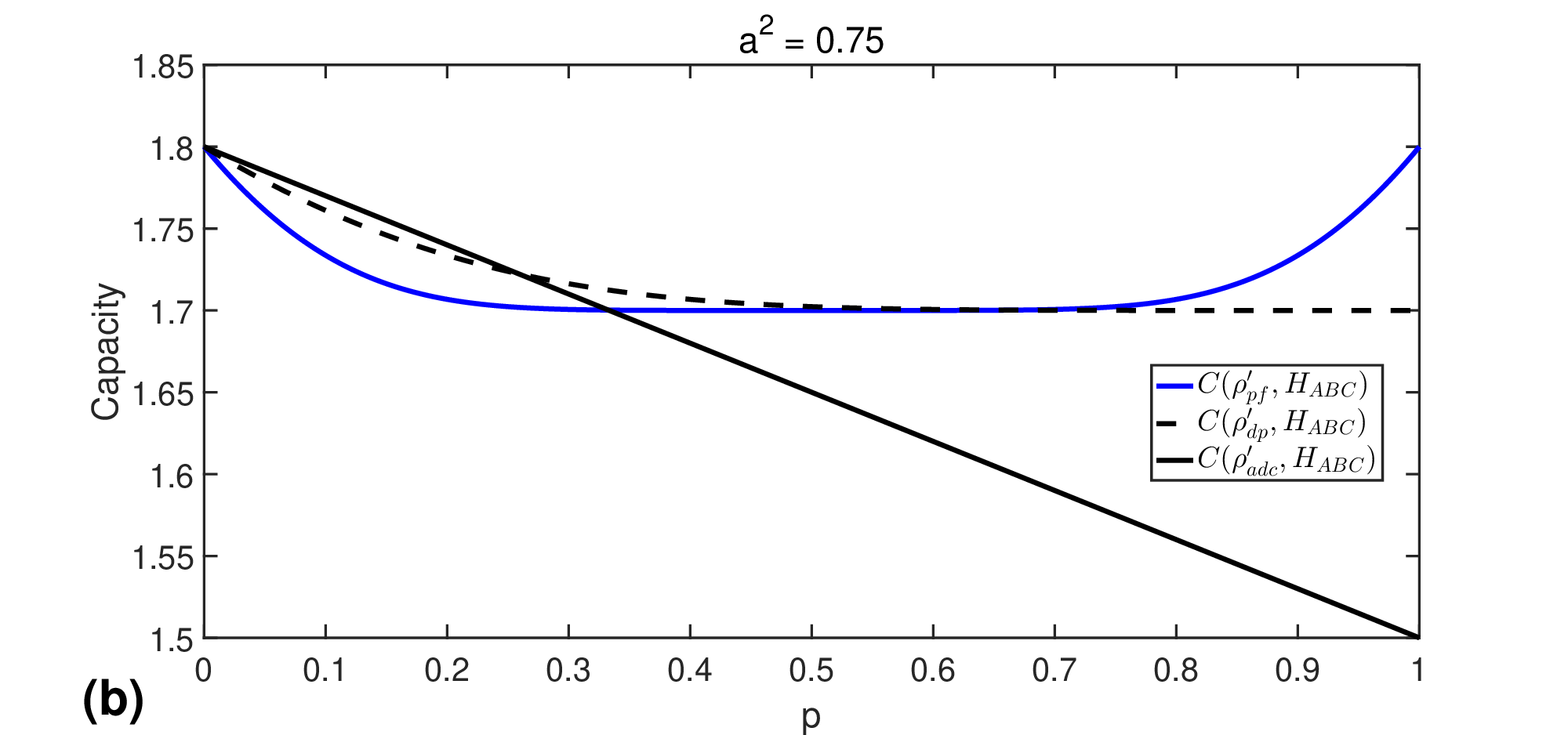}
			\label{Fig6b:right_mini}
		\end{minipage}
		\caption{ Quantum battery capacity evolution for GHZ-like states with $\epsilon^A=0.5$, $\epsilon^B=0.3$ and $\epsilon^c=0.1$ under the phase flip channel, dephasing channel and amplitude damping channel as a function of $p$. (a)$c_1=0.25=a^2$ (b) $c_1=0.75=a^2$}
		\label{Fig.6}
	\end{figure*}
	
	\subsection{B. Quantum battery capacity evolution under n times Markovian channels on the first system}
	Next, we consider the quantum battery capacity dynamics of GHZ-like states under n times Markovian channels. We find that phase flip channel and dephasing channel preserve the form of GHZ-like states.
	
	Let $\mathbi{C}(\rho_{pf}^{(n)},H_{ABC})$ and $\mathbi{C}(\rho_{dp}^{(n)},H_{ABC})$ be quantum battery capacity under n times the phase flip channel and n times the dephasing channel, respectively. $c_{1}^{'}, c_{2}^{'}, c_{3}^{'} \in[0,1]$ are given in Table IV.
	
	\begin{table}[htbp]
		\centering
		\begin{tabular}{|c|c|c|c|}
			\hline
			channel  & $c_{1}^{'}$  & $c_{2}^{'}$  & $c_{3}^{'}$  \\ \hline
			$pf^n$  & $c_1$  & $c_2(1-2p)^{3n}$  & $c_3$  \\ \hline
			$dp^n$  & $c_1$  & $c_2(1-p)^{3n}$  & $c_3$  \\ \hline
		\end{tabular}\\
		\vspace{4pt}
		{\footnotesize
			\textbf{Table IV:} Correlation coefficients for n times quantum channels: phase flip channel ($pf^n$) and dephasing channel ($dp^n$).
		}
	\end{table}
	
	From (\ref{e6}), we have
	\begin{equation*}
		\begin{split}
			&\mathbi{C}(\rho_{pf}^{(n)},H_{ABC})\\
			&=(c_1+c_3+\sqrt{(c_1-c_3)^2+4c_{2}^{2}(1-2p)^{6n}})\times 0.9\\
			&+(c_1+c_3-\sqrt{(c_1-c_3)^2+4c_{2}^{2}(1-2p)^{6n}})\times 0.7
		\end{split}
	\end{equation*}
	\begin{equation*}
		\begin{split}
			&\mathbi{C}(\rho_{dp}^{(n)},H_{ABC})\\
			&=(c_1+c_3+\sqrt{(c_1-c_3)^2+4c_{2}^{2}(1-p)^{6n}})\times 0.9\\
			&+(c_1+c_3+\sqrt{(c_1-c_3)^2+4c_{2}^{2}(1-p)^{6n}})\times 0.7.
		\end{split}
	\end{equation*}
	The detailed calculation process can be found in the Appendix E.
	Figure 7 (a) and (b) respectively show the evolution of the quantum battery capacity of GHZ-like states under n times phase flip channels and n times dephasing channels when $a^2=0.25$. It can be observed that as n increases, the rate of capacity decay of the battery accelerates, and the phenomenon of capacity freezing occurs earlier. The occurrence of the frozen capacity phenomenon indicates that the storage of battery capacity possesses a strong resistance to noise interference. When $a_2=c_1=0.75$, the evolution of quantum battery capacity for GHZ-like states is similar to Fig. 7 under the phase flip channel and the dephasing channel.
	\begin{figure*}[htbp]
		\centering
		\begin{minipage}[b]{0.48\linewidth}
			\centering
			\includegraphics[width=\linewidth]{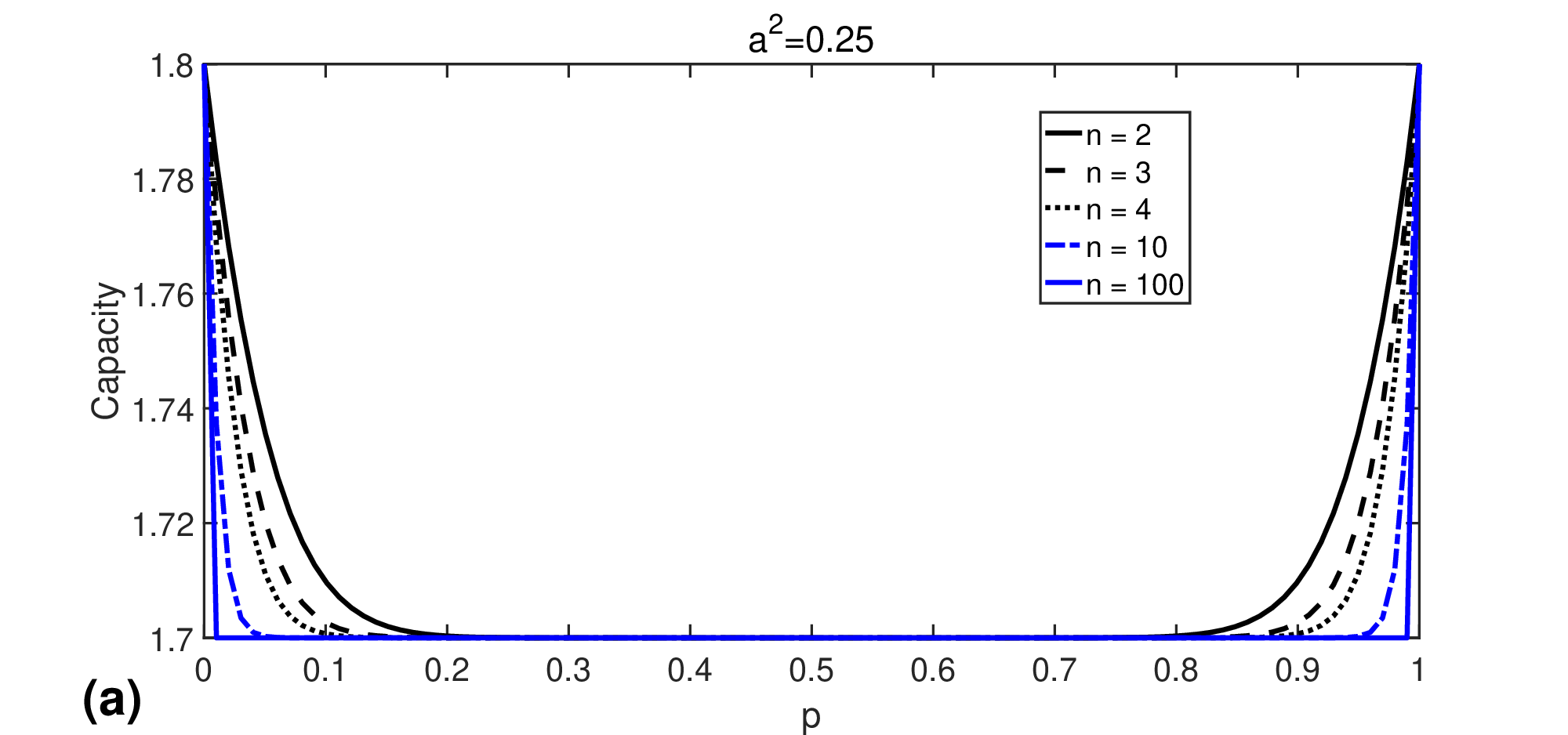}
			\label{Fig7a:left_mini}
		\end{minipage}
		\hfill
		\begin{minipage}[b]{0.48\linewidth}
			\centering
			\includegraphics[width=\linewidth]{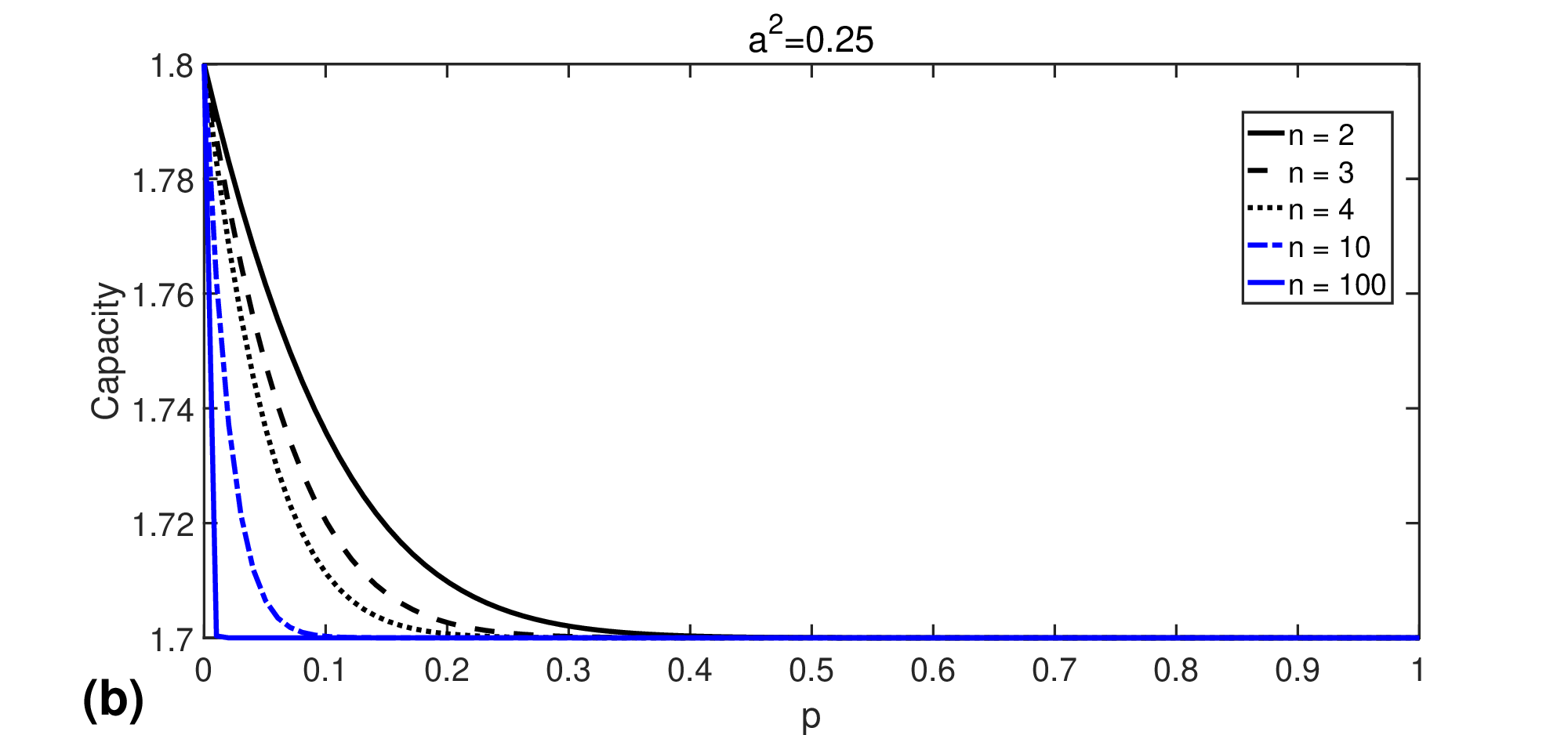}
			\label{Fig7b:right_mini}
		\end{minipage}
		\caption{(a) Quantum battery capacity evolution for GHZ-like state with $n=2,3,4,10,100$, $\epsilon^A=0.5$, $\epsilon^B=0.3$ and $\epsilon^c=0.1$ under the phase flip channel $n$ times, when $c_1=0.25=a^2$. (b) Quantum battery capacity evolution for GHZ-like state with n=2,3,4,10,100, $\epsilon^A=0.5$, $\epsilon^B=0.3$ and $\epsilon^c=0.1$ under the dephsaing channel $n$ times, when $c_1=0.25=a^2$.}
		\label{Fig.7}
	\end{figure*}
	
	Furthermore, after going through the amplitude damping channel n times on the first subsystem, we can obtain the capacity of the output state $\rho_{adc}^{(n)}$.
	 Using (\ref{e1}), when $c_1=a^2=0.25$, we have
	\begin{equation*}
		\begin{split}
			&\mathbi{C}(\rho_{adc}^{'},H_{ABC})\\
			&=\begin{cases}
				2\lambda_7(\epsilon^A+\epsilon^B+\epsilon^C)+2\lambda_5(\epsilon^A+\epsilon^B-\epsilon^C),  &p\in[0,x]\\
				2\lambda_5(\epsilon^A+\epsilon^B+\epsilon^C)+2\lambda_7(\epsilon^A+\epsilon^B-\epsilon^C),  &p\in(x,1],
				\end{cases}
		\end{split}
	\end{equation*}
	where, x is the intersection point of the eigenvalues $\lambda_5$ and $\lambda_7$. When $c_1=a^2=0.75$, we have
	$$
	\mathbi{C}(\rho_{adc}^{'},H_{ABC})
	=2\lambda_7(\epsilon^A+\epsilon^B+\epsilon^C)+2\lambda_5(\epsilon^A+\epsilon^B-\epsilon^C).
	$$
	The detailed calculation process can be found in the Appendix E3.
	Figure 8 shows the trend of quantum battery capacity under n times amplitude damping channels for GHZ-like states when $a^2=0.25$ and $a^2=0.75$. It is evident that the variation in quantum battery capacity is more stable when $a^2=0.75$ compared to $a^2=0.25$. Similarly, a capacity freezing phenomenon is also observed, and as n increases, the occurrence of capacity freezing happens earlier. This indicates that the quantum battery has a stronger capacity storage ability when $a^2=0.75$.
	
	\subsection{C. Quantum battery capacity dynamics of GHZ-like states under tri-side Markovian channels of the same type}
	In this part, we focus on the dynamics of quantum battery capacity of GHZ-like states undergoing three independent local Markovian channels of the same type, see the Kraus operators in Table II.
	The initial state $\rho$ evolves to another state $\rho^{'}=\varepsilon(\rho)$ under the Markovian channels. The corresponding coefficients are listed in the Table V.
	\begin{table}[htbp]
		\centering
		\begin{tabular}{|c|c|c|c|}
			\hline
			channel  & $c_{1}^{'}$  & $c_{2}^{'}$  & $c_{3}^{'}$  \\ \hline
			pf-pf-pf  & $c_1$  & $c_2(1-2p)(1-2q)(1-2\gamma)$  & $c_3$  \\ \hline
			dp-dp-dp  & $c_1$  & $c_2(1-p)(1-q)(1-\gamma)$  & $c_3$  \\ \hline
		\end{tabular}\\
		\vspace{4pt}
		{\footnotesize
			\textbf{Table V:} Correlation coefficients for  three independent local quantum channels: three independent local phase flip channel(pf-pf-pf) and three independent local dephasing channel(dp-dp-dp).
		}
	\end{table}
	
	From (\ref{e3}), we have the dynamical behaviours of the quantum battery capacity for GHZ-like states under the tri-side same type phase flip channel, see Fig. 9. The detailed calculation can be found in the Appendix E. Assuming that $\gamma$ is equal to 0.25, 0.5, 0.75, and 1, respectively. In figure 9 (a), as p and q increase, the capacity of the quantum battery first decreases and then increases, and when p = q = 0.5, the minimum value is reached. In figure 9 (b), regardless of the values of p and q, the quantum battery capacity is a constant. In figure 9 (c) and (d), the variation of the quantum battery capacity is similar to (a). Comparing the four sub-figures in Figure 9, we observe that when $\gamma = 0.5$, the quantum battery capacity remains constant regardless of the values of p and q. This indicates that under the tri-side same type phase flip channel, the quantum battery exhibits resistance to noise interference and possesses stable storage capabilities.
	\begin{figure*}[htbp]
	\centering
	\begin{minipage}[b]{0.48\linewidth}
		\centering
		\includegraphics[width=\linewidth]{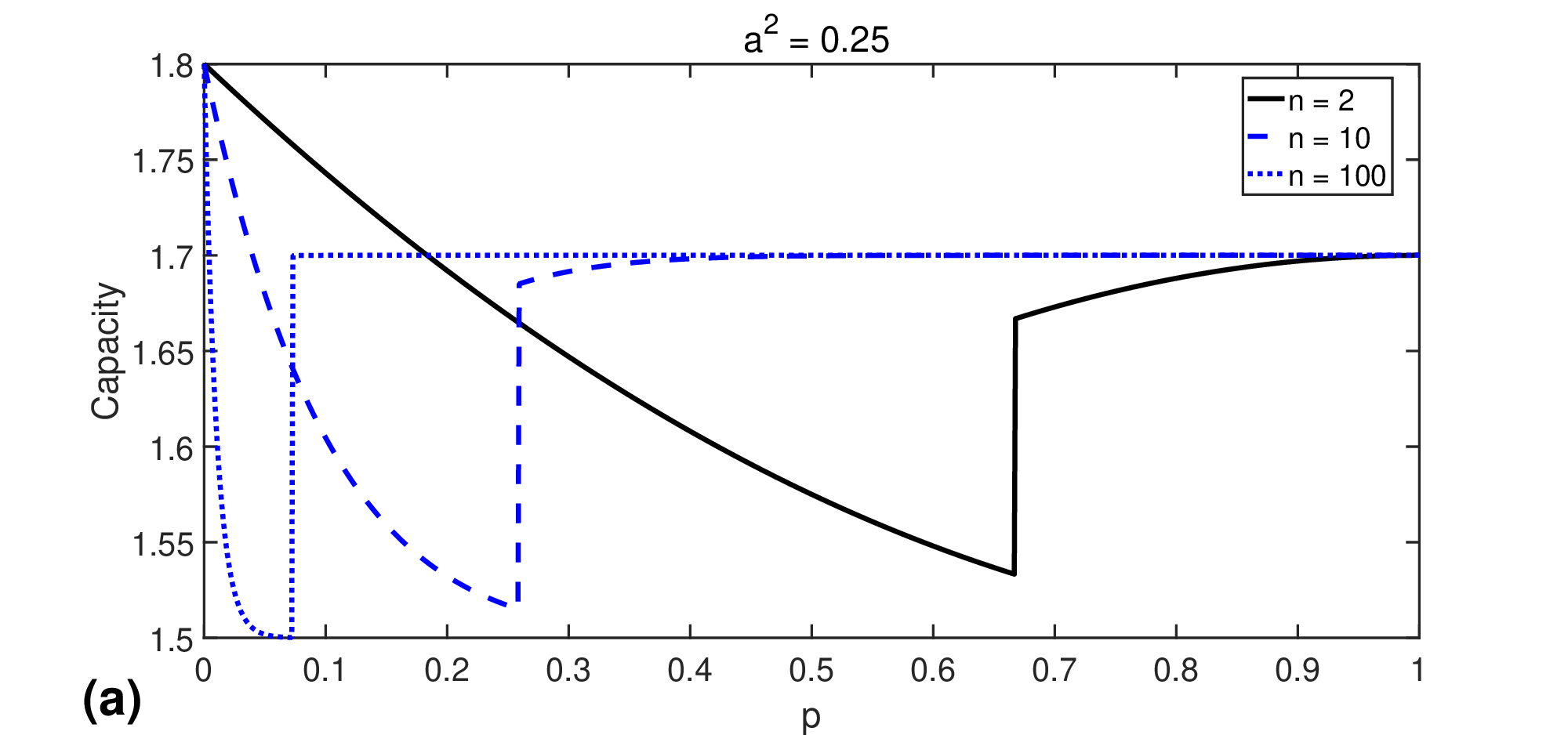}
		\label{Fig8a:left_mini}
	\end{minipage}
	\hfill
	\begin{minipage}[b]{0.48\linewidth}
		\centering
		\includegraphics[width=\linewidth]{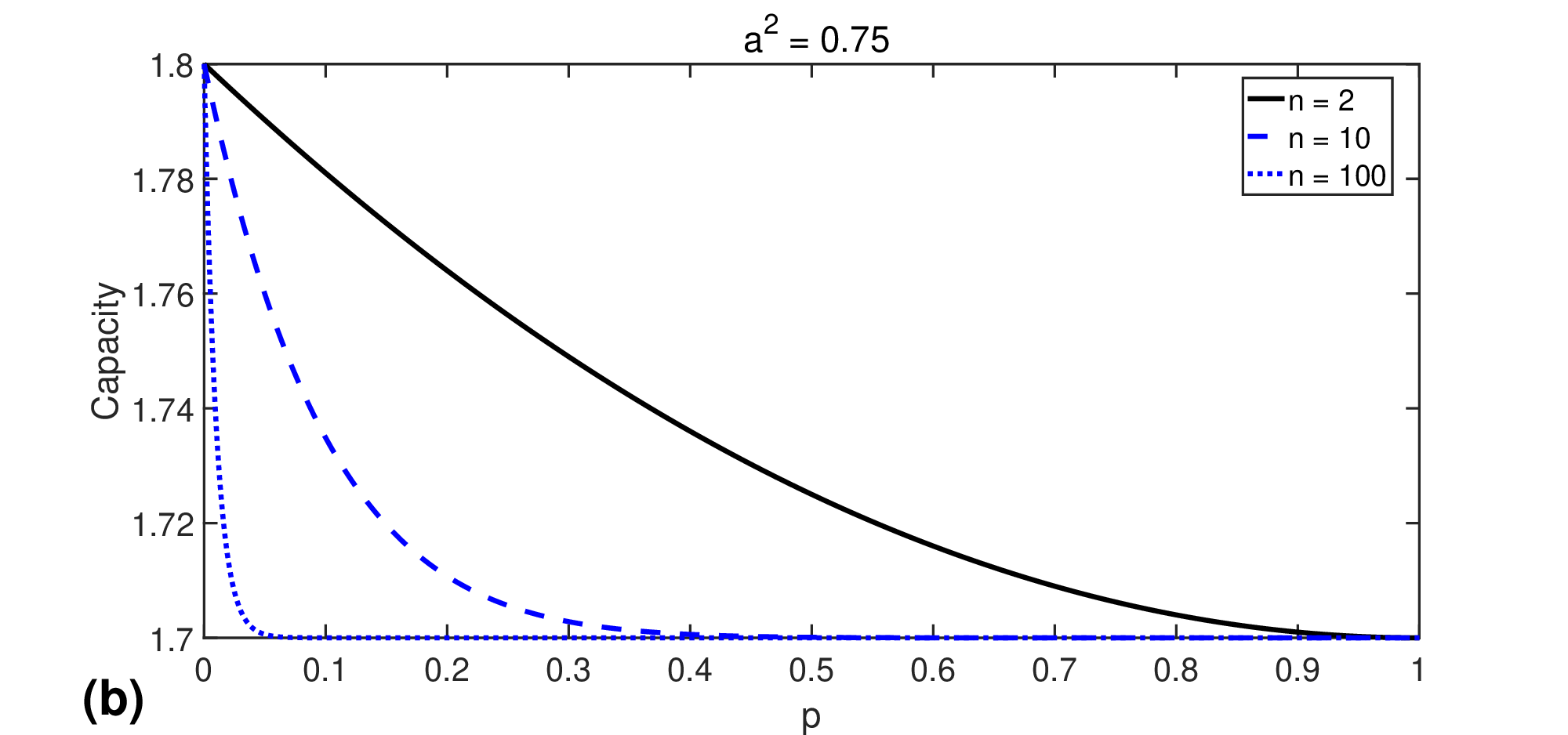}
		\label{Fig8b:right_mini}
	\end{minipage}
	\caption{Quantum battery capacity evolution for GHZ-like states with $n=2, 10, 100$, $\epsilon^A=0.5$, $\epsilon^B=0.3$ and $\epsilon^c=0.1$ under the amplitude damping channel $n$ times as a function of $p$. (a) $c_1=0.25=a^2$. (b) $c_1=0.75=a^2$.}
	\label{Fig.8}
\end{figure*}
	
	\begin{figure*}[htbp]
		\centering
		\begin{minipage}[b]{0.48\linewidth}
			\centering
			\includegraphics[width=\linewidth]{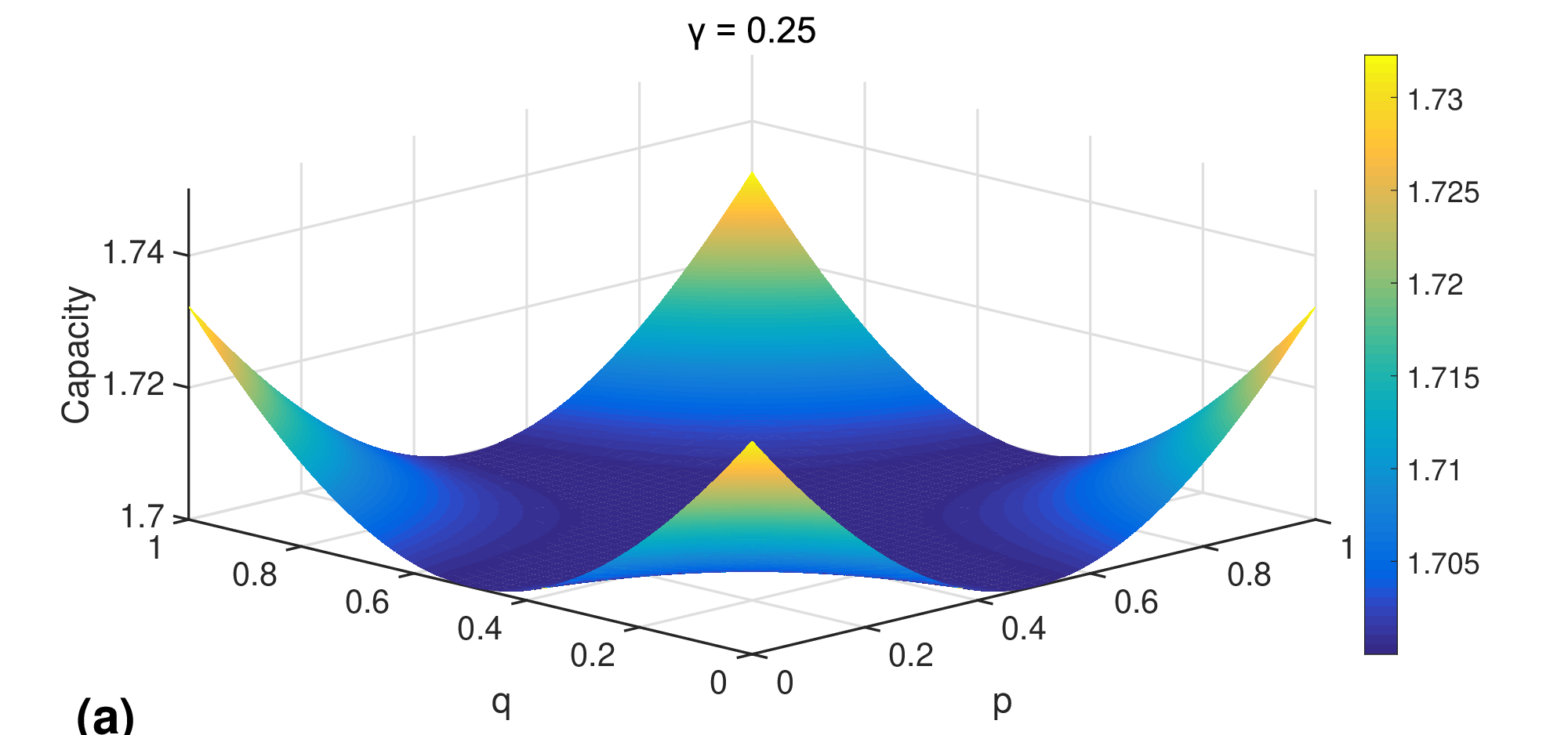}
			\label{Fig.9a}
		\end{minipage}
		\hfill
		\begin{minipage}[b]{0.48\linewidth}
			\centering
			\includegraphics[width=\linewidth]{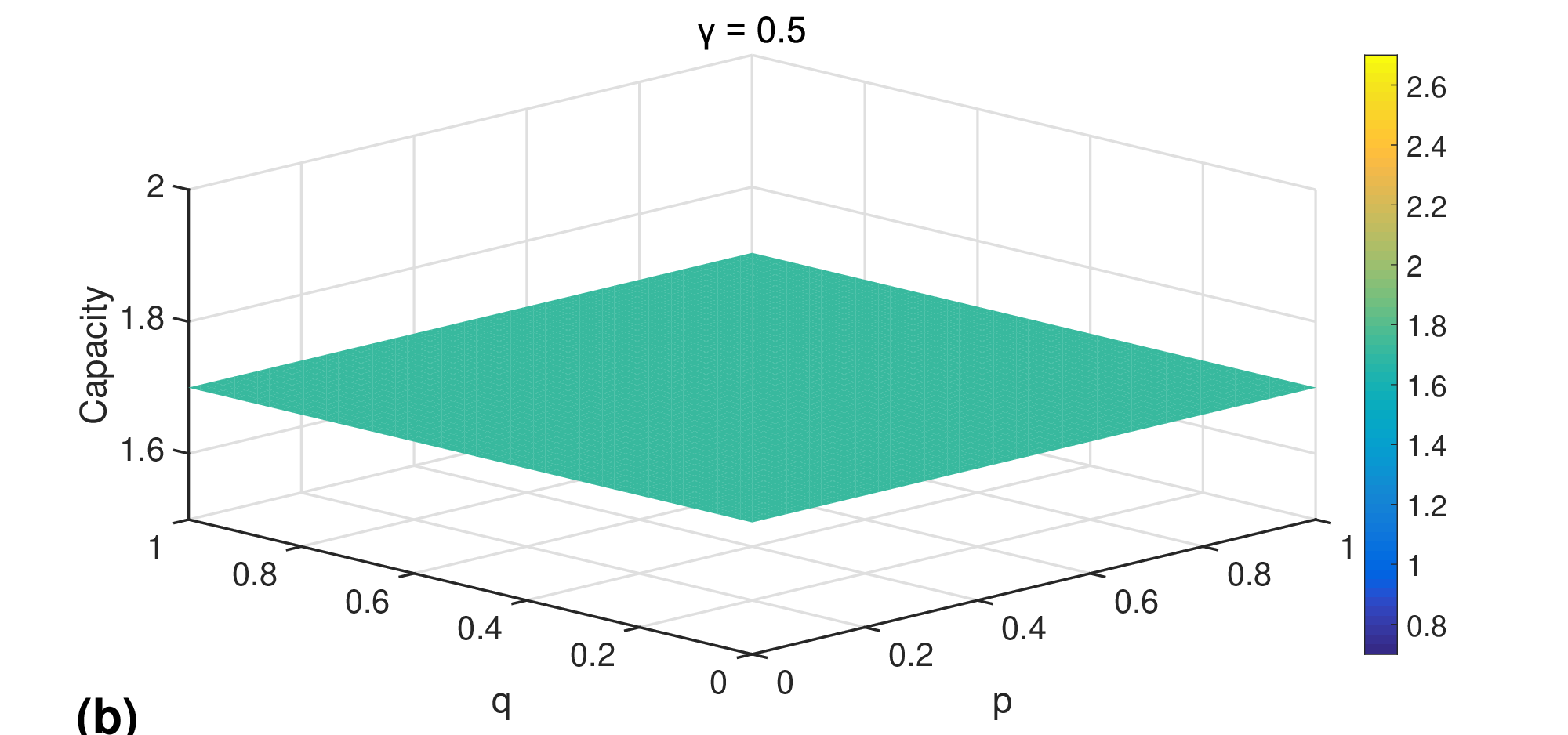}
			\label{Fig.9b}
		\end{minipage}
		\vspace{0.5em}
		\begin{minipage}[b]{0.48\linewidth}
			\centering
			\includegraphics[width=\linewidth]{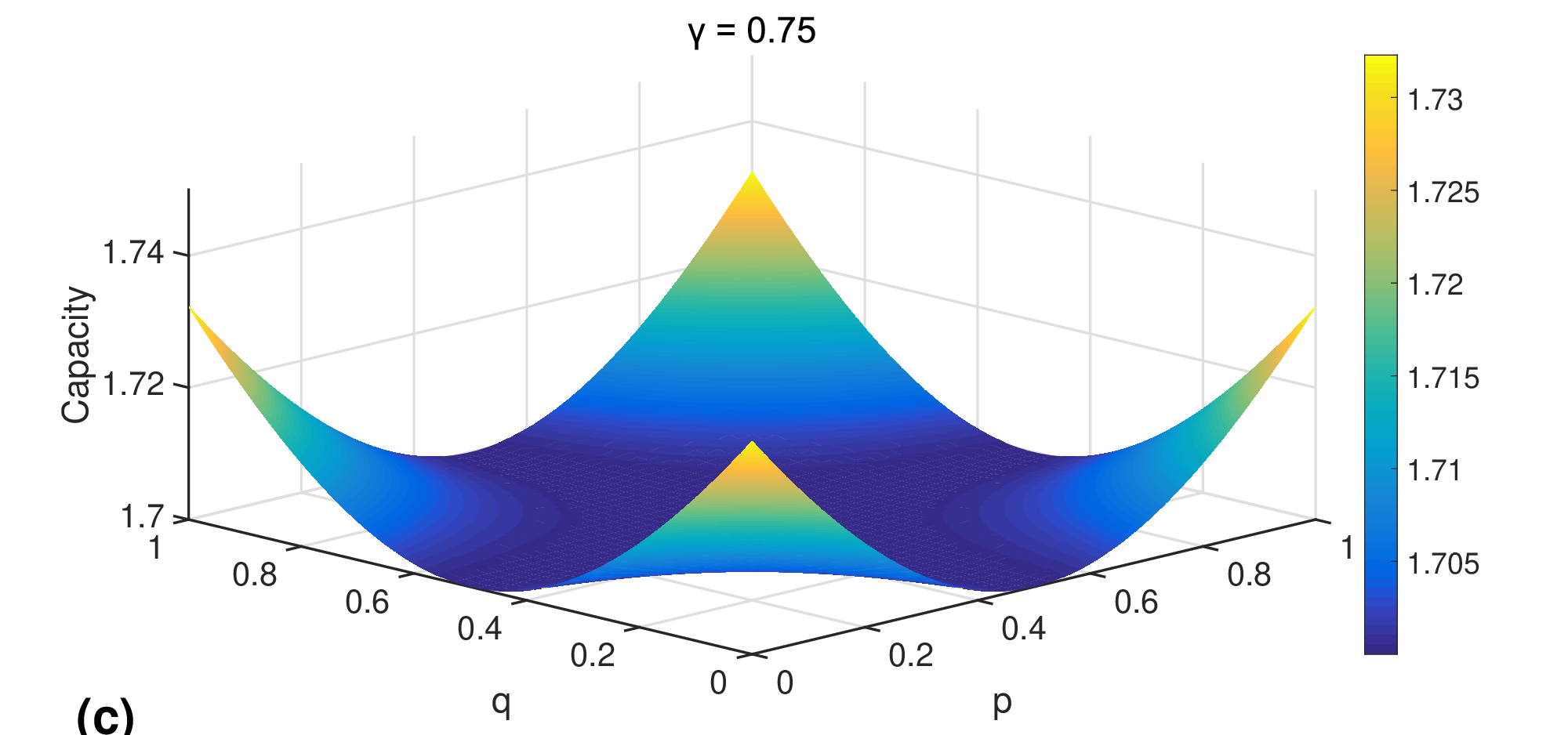}
			\label{Fig.9c}
		\end{minipage}
		\hfill
		\begin{minipage}[b]{0.48\linewidth}
			\centering
			\includegraphics[width=\linewidth]{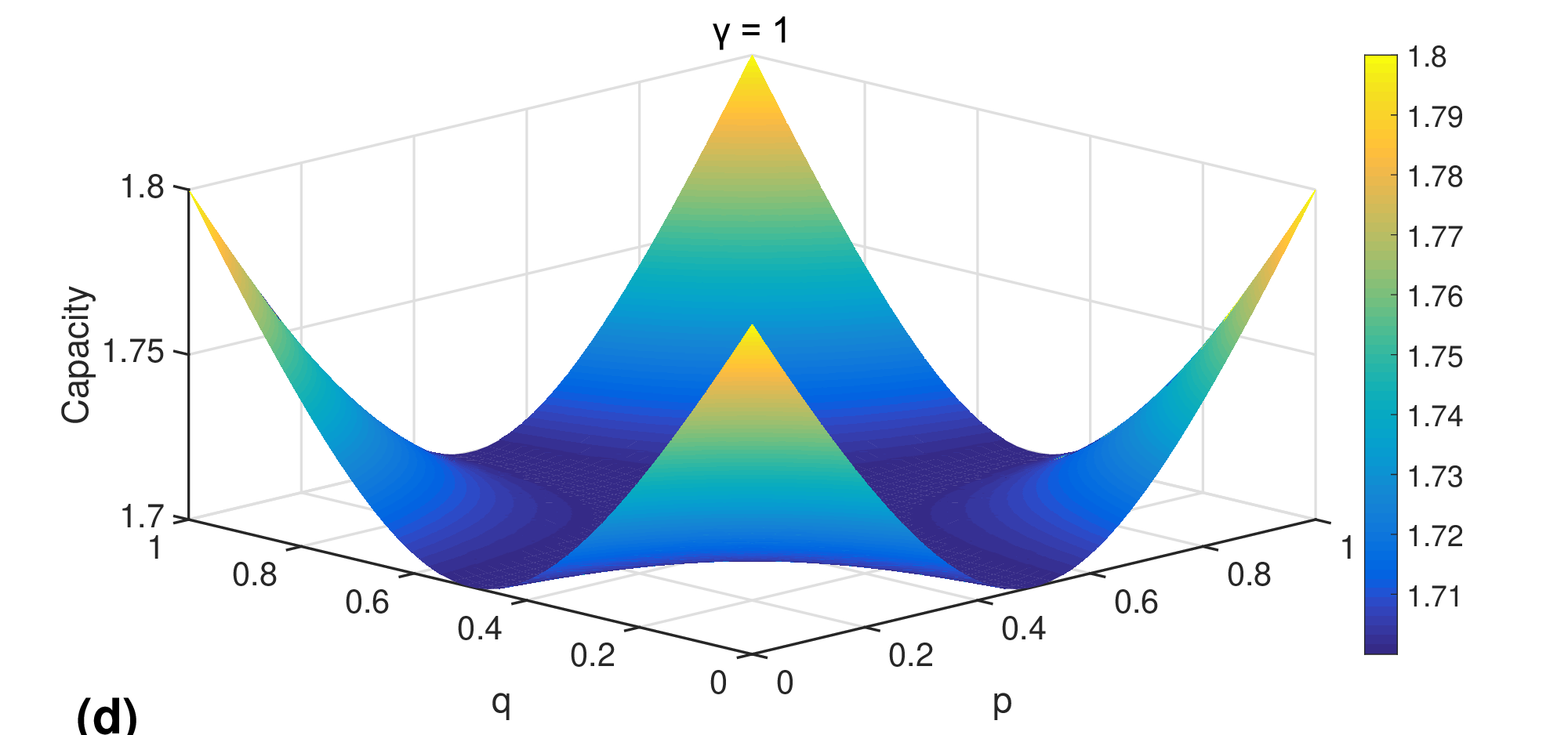}
			\label{Fig.9d}
		\end{minipage}
		
		\caption{ When $c_1=0.25=a^2$, quantum battery capacity evolution for GHZ-like states with $\epsilon^A=0.5$, $\epsilon^B=0.3$ and $\epsilon^C=0.1$ under tri-side phase flip channel. (a) $\gamma=0.25$  (b) $\gamma=0.5$ (c) $\gamma=0.75$ (d) $\gamma=1$
		}
		\label{Fig.9}
	\end{figure*}
	
	\begin{figure*}[htbp]
		\centering
		\begin{minipage}[b]{0.48\linewidth}
			\centering
			\includegraphics[width=\linewidth]{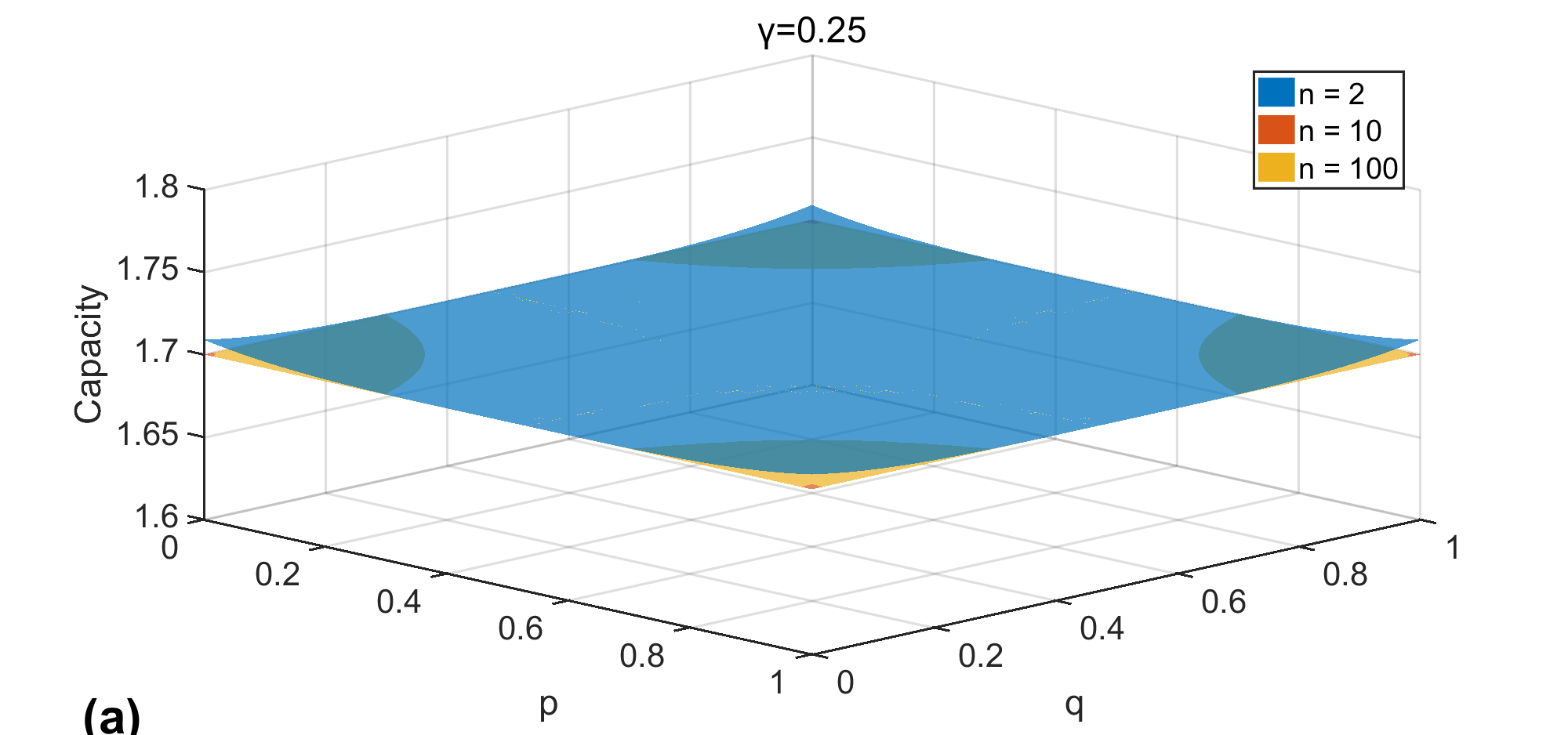}
			\label{Fig.10a}
		\end{minipage}
		\hfill
		\begin{minipage}[b]{0.48\linewidth}
			\centering
			\includegraphics[width=\linewidth]{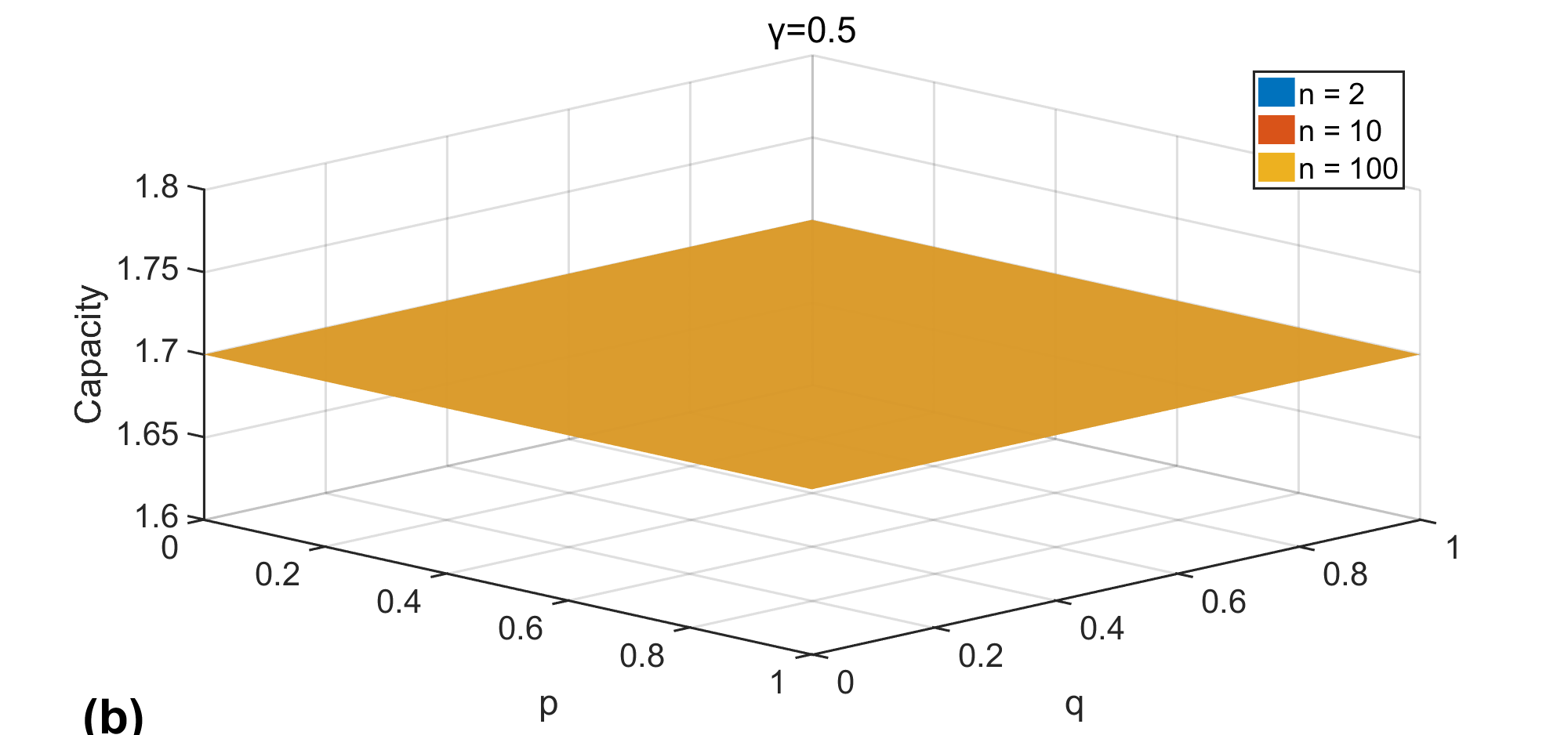}
			\label{Fig.10b}
		\end{minipage}
		\vspace{0.5em}
		\begin{minipage}[b]{0.48\linewidth}
			\centering
			\includegraphics[width=\linewidth]{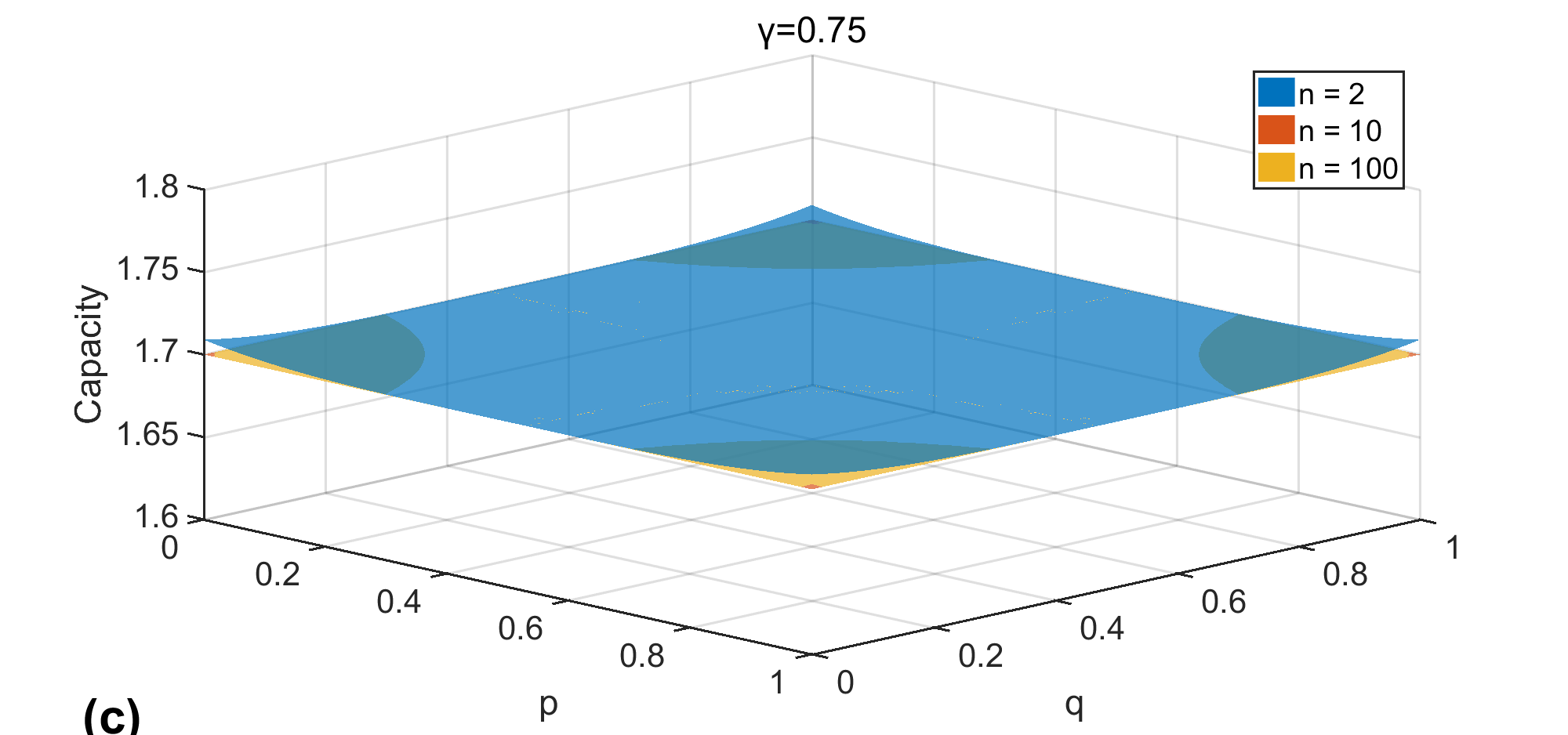}
			\label{Fig.10c}
		\end{minipage}
		\hfill
		\begin{minipage}[b]{0.48\linewidth}
			\centering
			\includegraphics[width=\linewidth]{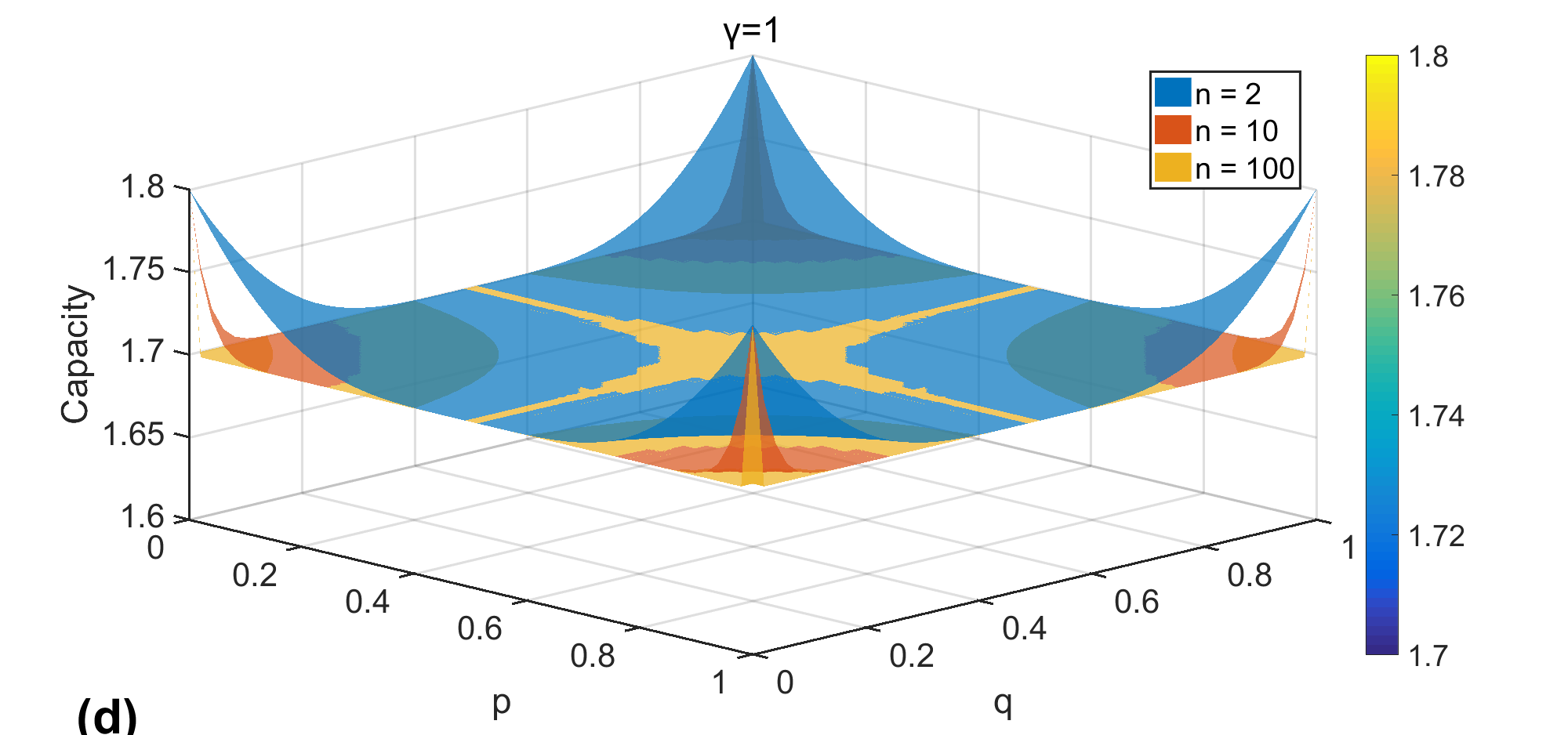}
			\label{Fig.10d}
		\end{minipage}
		
		\caption{ When $c_1=0.25=a^2$, quantum battery capacity evolution for GHZ-like states with $n=2, 10, 100$ and $\epsilon^A=0.5$, $\epsilon^B=0.3$, $\epsilon^C=0.1$ under tri-side phase flip channel $n$ times. (a) $\gamma=0.25$ (b) $\gamma=0.5$ (c) $\gamma=0.75$ (d) $\gamma=1$
		}
		\label{Fig.10}
	\end{figure*}
	Furthermore, if all subsystems go through the same type channel n times, the initial state $\rho_{ABC}$  also evolves to another state $\rho^{'}=\varepsilon(\rho)$ under the Markovian channels. The corresponding coefficients are listed in the Table VI.
	\begin{table}[htbp]
		\centering
		\begin{tabular}{|c|c|c|c|}
			\hline
		channel  & $c_{1}^{'}$  & $c_{2}^{'}$  & $c_{3}^{'}$  \\ \hline
			$pf^n-pf^n-pf^n$  & $c_1$  & $c_2(1-2p)^n(1-2q)^n(1-2\gamma)^n$  & $c_3$  \\ \hline
			$dp^n-dp^n-dp^n$  & $c_1$  & $c_2(1-p)^n(1-q)^n(1-\gamma)^n$  & $c_3$  \\ \hline
		\end{tabular}\\
		\vspace{4pt}
		{\footnotesize
			\textbf{Table VI:} Correlation coefficients for n times quantum operations: three independent local phase flip channel($pf^n-pf^n-pf^n$) and three independent local dephasing channel($dp^n-dp^n-dp^n$).
		}
	\end{table}
	By (\ref{e3}), we can observe the dynamical behaviours of the quantum battery capacity for GHZ-like states under the tri-side same type phase flip channel n times, see Fig. 10. Similarly, the detailed calculation can be found in the Appendix E1. Assuming that $\gamma$ is equal to $0.25, 0.5, 0.75$, and $1$, respectively. In figure 10 (a), as $p$ and $q$ increase, the quantum battery capacity first decreases and then increases. And the larger n, the faster the quantum battery capacity decreases. In figure 10 (b), regardless of the values of $p$ and $q$, the quantum battery capacity is a constant. In figure 10 (c) and (d), the quantum battery capacity is similar to (a). Comparing the four sub-figures in Figure 10, we observe that when$\gamma = 0.5$, regardless of the values of p and q or the value of n, the quantum battery capacity remains constant. This indicates that under n times tri-side same type phase flip channels, the quantum battery exhibits stronger resilience to noise interference and enhanced storage capacity.

	\section{IV. SUMMARY}
	In this work, we mainly study the evolution of quantum battery capacity of GHZ state and GHZ-like states under Markovian channels in the tripartite system. The GHZ state occurs frozen capacity phenomenon under the dephasing channel; there may be a brief sudden death phenomenon of quantum battery capacity under bit-phase flip channel and depolarizing channel; under the amplitude damping channel and on the first subsystem, the quantum battery capacity monotonically decreases. In the presence of n times dephasing channels and amplitude damping channels, the battery capacity declines to a fixed value, known as the frozen capacity, and this phenomenon occurs earlier as n increases. In the case of the tri-side same type phase flip channel, the quantum battery capacity remains at a relatively high level. For GHZ-like states, the variations in battery capacity are studied for parameters $c_1 = a^2 = 0.25$ and $c_2 = a^2 = 0.75$. Taking $c_1 = a^2 = 0.25$ as an example, the GHZ-like state exhibits a phenomenon of capacity freezing under the dephasing channel; under n times dephasing channel, the battery capacity shows freezing behaviour, and the larger the n, the earlier the freezing phenomenon occurs. In the case of a three-sided identical phase-flip channel, even after experiencing multiple noise interference, the quantum battery can still maintain a certain energy storage capacity. 
	
    We can observe that the quantum battery capacity storage capability of the GHZ state is stronger under amplitude damping channels, demonstrating superior performance. In contrast, the GHZ-like state exhibits higher stability of battery capacity under dephasing channels, indicating that different noise channels have varying impacts on different initial states. Compared to the Bell-diagonal state in a bipartite systems, the GHZ state does not enhance the quantum battery capacity after the action of a Markovian channel. Since all results are analytical and based on widely recognized noise models, they provide clear and testable predictions for experimental physicists. As the maximally entangled state, the GHZ state is neither an overly simplified subject that loses practical significance nor an abstraction that is too complex to be analytically understood. It can clearly reveal the key features and fundamental laws of quantum battery capacity evolution in noisy environments. This model serves as an ideal starting point and benchmark; after fully understanding the behavior of the tripartite system under local noise, researchers can gradually increase complexity based on this foundation. In the evolution of quantum battery capacity under phase-flip channels, non-Markovianity emerges, indicating that the capacity can recover. This phenomenon may potentially be leveraged in future research to achieve higher charging power, greater stable energy, and more reliable storage capacity. This study reveals phenomena in noisy environments such as brief sudden death and frozen capacity by analyzing the impact of different Markovian channels on the capacity of quantum batteries, providing a theoretical basis for the practical application of quantum energy storage technology.

\begin{acknowledgments}
This work is supported by the National Natural Science Foundation of China (NSFC) under Grant Nos. 12204137 and 12564048; the Natural Science Foundation of Hainan Province under Grant No. 125RC744 and the Hainan Academician Workstation.
\end{acknowledgments}

\onecolumngrid
\appendix

\section{Appendix A: The calculation process of the Eq.(\ref{e3})}

Consider the three-qubit GHZ state,
\begin{equation*}
	\begin{split}
		\rho_{ABC}=|GHZ\rangle \langle GHZ|
		=\frac{1}{2}(|000 \rangle+|111\rangle)(\langle 000|+\langle 111|).
	\end{split}
\end{equation*}
where $|GHZ\rangle =\frac{1}{\sqrt 2}(|000\rangle +|111\rangle)$.
The eigenvalues of $ \rho_{ABC} $ are $\lambda_i=0,(i=0,...,6)$, $\lambda_7=1$. The Hamiltonian of the tripartite system is 
$$ 
H_{ABC}=\epsilon^A \sigma_3 \otimes I_2 \otimes I_2+ \epsilon^BI_2 \otimes \sigma_3 \otimes I_2+ \epsilon^C I_2 \otimes I_2 \otimes \sigma_3,
$$ 
where $\sigma_3$ is the standard Pauli matrices, $I_2$ is the identity operator, and $0 \leqslant \epsilon^C \leqslant \epsilon^B \leqslant \epsilon^A $. The eigenvalues of $ H_{ABC} $ are $ -\epsilon^A-\epsilon^B-\epsilon^c, -\epsilon^A-\epsilon^B+\epsilon^c, -\epsilon^A+\epsilon^B-\epsilon^c, -\epsilon^A+\epsilon^B+\epsilon^c, \epsilon^A-\epsilon^B-\epsilon^c, \epsilon^A-\epsilon^B+\epsilon^c, \epsilon^A+\epsilon^B-\epsilon^c, \epsilon^A+\epsilon^B+\epsilon^c $, and arranged in ascending order. Therefore, from (\ref{e1}) we have
\begin{equation*}
	\begin{split}
		\mathbi{C}(\rho_{ABC},H_{ABC})&=\epsilon_0(\lambda_0-\lambda_7)+\epsilon_1(\lambda_1-\lambda_6)+\epsilon_2(\lambda_2-\lambda_5+\epsilon_3(\lambda_3-\lambda_4)+\epsilon_4(\lambda_4-\lambda_3)+\epsilon_5(\lambda_5-\lambda_2)\\
		&+\epsilon_6(\lambda_6-\lambda_1)+\epsilon_7(\lambda_7-\lambda_0)\\
		&=2\epsilon_4(\lambda_4-\lambda_3)+2\epsilon_5(\lambda_5-\lambda_2)+2\epsilon_6(\lambda_6-\lambda_1)+2\epsilon_7(\lambda_7-\lambda_0)\\
		&=2(\epsilon^A+\epsilon^B+\epsilon^c).
	\end{split}
\end{equation*}

\section{Appendix B: Quantum battery capacity of GHZ state under Markovian channels.}
\subsection{B1. Under the bit flip channel}

From (\ref{e4}), we have 
\begin{equation*}
	\begin{split}
		\rho_{bf}^{'}=\varepsilon(\rho)=E_{000}\rho E_{000}^{\dagger}+E_{001}\rho E_{001}^{\dagger}+E_{010}\rho E_{010}^{\dagger}+E_{100}\rho E_{100}^{\dagger}+E_{011}\rho E_{011}^{\dagger}+E_{101}\rho E_{101}^{\dagger}+E_{110}\rho E_{110}^{\dagger}+E_{111}\rho E_{111}^{\dagger}.
	\end{split}
\end{equation*}
The table I lists the Kraus operations, so we have the output state $\rho_{bf}^{'}$ under the bit flip channel,
 \begin{equation*}
 		\begin{split}
 			\rho_{bf}^{'}=\frac{1}{2}\left(\begin{array}{cccccccc}
 				1-\frac{3p}{2}+\frac{3p^2}{4} & 0 & 0 & 0 & 0 & 0 & 0 & 1-\frac{3p}{2}+\frac{3p^2}{4} \\
 				0 & \frac{p}{2}-\frac{p^2}{4} & 0 & 0 & 0 & 0 & \frac{p}{2}-\frac{p^2}{4} & 0 \\
 				0 & 0 & \frac{p}{2}-\frac{p^2}{4} & 0 & 0 & \frac{p}{2}-\frac{p^2}{4} & 0 & 0 \\
 				0 & 0 & 0 & \frac{p}{2}-\frac{p^2}{4} & \frac{p}{2}-\frac{p^2}{4} & 0 & 0 & 0 \\
 				0 & 0 & 0 & \frac{p}{2}-\frac{p^2}{4} & \frac{p}{2}-\frac{p^2}{4} & 0 & 0 & 0 \\
 				0 & 0 &\frac{p}{2}-\frac{p^2}{4} & 0 & 0 & \frac{p}{2}-\frac{p^2}{4} & 0 & 0\\
 				0 & \frac{p}{2}-\frac{p^2}{4} & 0 & 0 & 0 & 0 & \frac{p}{2}-\frac{p^2}{4} & 0 \\
 				1-\frac{3p}{2}+\frac{3p^2}{4} & 0 & 0 & 0 & 0 & 0 & 0 & 1-\frac{3p}{2}+\frac{3p^2}{4}
 			\end{array}\right)
       \end{split}
 	\end{equation*}
So the eigenvalues of $\rho_{bf}^{'}$ are $\lambda_i=0,(i=0,...,3)$,$\lambda_4=\lambda_5=\lambda_6=p-p^2$ and $\lambda_7=1-3p+3p^2$. From (\ref{e1}), we have the quantum battery capacity
\begin{equation*}
	\begin{split}
		\mathcal{C}(\rho_{bf}^{'};H)&=2(\lambda_7-\lambda_0)(\epsilon^A+\epsilon^B+\epsilon^C)+2(\lambda_6-\lambda_1)(\epsilon^A+\epsilon^B-\epsilon^C)+2(\lambda_5-\lambda_2)(\epsilon^A-\epsilon^B+\epsilon^C)+2(\lambda_4-\lambda_3)(\epsilon^A-\epsilon^B-\epsilon^C)\\
		&=2(1-3p+3p^2)(\epsilon^A+\epsilon^B+\epsilon^C)+2(p-p^2)(3\epsilon^A-\epsilon^B-\epsilon^C)	
	\end{split}
\end{equation*}
\subsection{B2. Under the dephasing channel}

Similarly, in this situation, we have
\begin{equation*}
	\begin{split}
		\rho_{dp}^{'}=\varepsilon(\rho)=E_{000}\rho E_{000}^{\dagger}+E_{001}\rho E_{001}^{\dagger}+E_{010}\rho E_{010}^{\dagger}+E_{100}\rho E_{100}^{\dagger}+E_{011}\rho E_{011}^{\dagger}+E_{101}\rho E_{101}^{\dagger}+E_{110}\rho E_{110}^{\dagger}+E_{111}\rho E_{111}^{\dagger}.
	\end{split}
\end{equation*}
So we have 
\begin{equation*}
		\begin{split}
			\rho_{dp}^{'}=\frac{1}{2}\left(\begin{array}{cccccccc}
				1 & 0 & 0 & 0 & 0 & 0 & 0 & (1-p)^3 \\
				0 & 0 & 0 & 0 & 0 & 0 & 0 & 0 \\
				0 & 0 & 0 & 0 & 0 & 0 & 0 & 0 \\
				0 & 0 & 0 & 0 & 0 & 0 & 0 & 0 \\
				0 & 0 & 0 & 0 & 0 & 0 & 0 & 0 \\
				0 & 0 & 0 & 0 & 0 & 0 & 0 & 0 \\
				0 & 0 & 0 & 0 & 0 & 0 & 0 & 0 \\
				(1-p)^3 & 0 & 0 & 0 & 0 & 0 & 0 & 1
			\end{array}\right)
		\end{split}
\end{equation*}
Therefore the eigenvalues of $\rho_{dp}^{'}$ are $\lambda_i=0, (i=1,...,5)$, $\lambda_6=\frac{3p}{2}-\frac{3p^2}{2}+\frac{p^3}{2}$ and $\lambda_7=1-\frac{3p}{2}+\frac{3p^2}{2}-\frac{p^3}{2}$. According to (\ref{e1}), we also have capacity
\begin{equation*}
	\begin{split}
		\mathcal{C}(\rho_{dp}^{'};H)&=2(\lambda_7-\lambda_0)(\epsilon^A+\epsilon^B+\epsilon^C)+2(\lambda_6-\lambda_1)(\epsilon^A+\epsilon^B-\epsilon^C)+2(\lambda_5-\lambda_2)(\epsilon^A-\epsilon^B+\epsilon^C)+2(\lambda_4-\lambda_3)(\epsilon^A-\epsilon^B-\epsilon^C)\\
		&=(2-3p+3p^2-p^3)(\epsilon^A+\epsilon^B+\epsilon^C)+(3p-3p^2+p^3)(\epsilon^A+\epsilon^B-\epsilon^C)	
	\end{split}
\end{equation*}

\subsection{B3.Under the phase flip channel}

According to (\ref{e4}), after the trace-preserving quantum operation, the output state is
\begin{equation*}
	\begin{split}
		\rho_{pf}^{'}=\varepsilon(\rho)=E_{000}\rho E_{000}^{\dagger}+E_{001}\rho E_{001}^{\dagger}+E_{010}\rho E_{010}^{\dagger}+E_{100}\rho E_{100}^{\dagger}+E_{011}\rho E_{011}^{\dagger}+E_{101}\rho E_{101}^{\dagger}+E_{110}\rho E_{110}^{\dagger}+E_{111}\rho E_{111}^{\dagger}.
	\end{split}
\end{equation*}
And
\begin{equation*}
	\begin{split}
		\rho_{pf}^{'}=\frac{1}{2}\left(\begin{array}{cccccccc}
			1 & 0 & 0 & 0 & 0 & 0 & 0 & (1-2p)^3 \\
			0 & 0 & 0 & 0 & 0 & 0 & 0 & 0 \\
			0 & 0 & 0 & 0 & 0 & 0 & 0 & 0 \\
			0 & 0 & 0 & 0 & 0 & 0 & 0 & 0 \\
			0 & 0 & 0 & 0 & 0 & 0 & 0 & 0 \\
			0 & 0 & 0 & 0 & 0 & 0 & 0 & 0 \\
			0 & 0 & 0 & 0 & 0 & 0 & 0 & 0 \\
			(1-2p)^3 & 0 & 0 & 0 & 0 & 0 & 0 & 1
		\end{array}\right)
	\end{split}
\end{equation*}
So the eigenvalues of $\rho_{pf}^{'}$ are $\lambda_i=0,(i=0,...,5)$, $\lambda_6=3p-6p^2+4p^3$ and $\lambda_7=1-3p+6p^2-4p^3$.
When $0\leqslant p\leqslant\frac{1}{2}$, the quantum battery capacity is
\begin{equation*}
	\begin{split}
		\mathcal{C}(\rho_{pf}^{'};H)&=2(\lambda_7-\lambda_0)(\epsilon^A+\epsilon^B+\epsilon^C)+2(\lambda_6-\lambda_1)(\epsilon^A+\epsilon^B-\epsilon^C)+2(\lambda_5-\lambda_2)(\epsilon^A-\epsilon^B+\epsilon^C)+2(\lambda_4-\lambda_3)(\epsilon^A-\epsilon^B-\epsilon^C)\\
		&=2(1-3p+6p^2-4p^3)(\epsilon^A+\epsilon^B+\epsilon^C)+2(3p-6p^2+4p^3)(\epsilon^A+\epsilon^B-\epsilon^C)	
	\end{split}
\end{equation*}
When $\frac{1}{2} < p\leqslant 1$, we have
\begin{equation*}
	\begin{split}
		\mathcal{C}(\rho_{pf}^{'};H)&=2(\lambda_7-\lambda_0)(\epsilon^A+\epsilon^B+\epsilon^C)+2(\lambda_6-\lambda_1)(\epsilon^A+\epsilon^B-\epsilon^C)+2(\lambda_5-\lambda_2)(\epsilon^A-\epsilon^B+\epsilon^C)+2(\lambda_4-\lambda_3)(\epsilon^A-\epsilon^B-\epsilon^C)\\
		&=2(3p-6p^2+4p^3)(\epsilon^A+\epsilon^B+\epsilon^C)+2(1-3p+6p^2-4p^3)(\epsilon^A+\epsilon^B-\epsilon^C)	.
	\end{split}
\end{equation*}
Therefore, the quantum battery capacity is
\begin{equation*}
	\mathbi{C}(\rho_{pf}^{'},H_{ABC})=\begin{cases}
		2(1-3p+6p^2-4p^3)(\epsilon^A+\epsilon^B+\epsilon^C)+2(3p-6p^2+4p^3)(\epsilon^A+\epsilon^B-\epsilon^C),&p\in[0,\frac{1}{2}]\\
		2(3p-6p^2+4p^)(\epsilon^A+\epsilon^B+\epsilon^C)+2(1-3p+6p^2-4p^3)(\epsilon^A+\epsilon^B-\epsilon^C),&p\in(\frac{1}{2},1]
	\end{cases}
\end{equation*}

\subsection{B4. Under the depolarizing channel}

From (\ref{e4}), we obtain the output state
	\begin{equation*}
	\begin{split}
		\rho_{dep}^{'}=\varepsilon(\rho)=E_{000}\rho E_{000}^{\dagger}+E_{001}\rho E_{001}^{\dagger}+E_{010}\rho E_{010}^{\dagger}+E_{100}\rho E_{100}^{\dagger}+E_{011}\rho E_{011}^{\dagger}+E_{101}\rho E_{101}^{\dagger}+E_{110}\rho E_{110}^{\dagger}+E_{111}\rho E_{111}^{\dagger}.
	\end{split}
\end{equation*}
And
\begin{equation*}
	\begin{split}
		\rho_{dep}^{'}=\frac{1}{2}\left(\begin{array}{cccccccc}
			1-2p+\frac{4p^2}{3} & 0 & 0 & 0 & 0 & 0 & 0 & (1-\frac{4p}{3})^3 \\
			0 & \frac{2p}{3}-\frac{4p^2}{9} & 0 & 0 & 0 & 0 & 0 & 0 \\
			0 & 0 & \frac{2p}{3}-\frac{4p^2}{9} & 0 & 0 & 0 & 0 & 0 \\
			0 & 0 & 0 & \frac{2p}{3}-\frac{4p^2}{9} & 0 & 0 & 0 & 0 \\
			0 & 0 & 0 & 0 & \frac{2p}{3}-\frac{4p^2}{9} & 0 & 0 & 0 \\
			0 & 0 & 0 & 0 & 0 & \frac{2p}{3}-\frac{4p^2}{9} & 0 & 0 \\
			0 & 0 & 0 & 0 & 0 & 0 & \frac{2p}{3}-\frac{4p^2}{9} & 0 \\
			(1-\frac{4p}{3})^3 & 0 & 0 & 0 & 0 & 0 & 0 & 1-2p+\frac{4p^2}{3}
		\end{array}\right)
	\end{split}
\end{equation*}
So the eigenvalues of $\rho_{dep}^{'}$ are $\lambda_i=\frac{p}{3}-\frac{2p^2}{9},(i=1,...,5)$, $\lambda_6=p-2p^2+\frac{32p^3}{27}$ and$\lambda_7=1-3p+\frac{10p^2}{3}-\frac{32p^3}{27}$. When$0\leqslant p\leqslant \frac{3}{4}$,
\begin{equation*}
	\begin{split}
		\mathcal{C}(\rho_{dep}^{'};H)&=2(\lambda_7-\lambda_0)(\epsilon^A+\epsilon^B+\epsilon^C)+2(\lambda_6-\lambda_1)(\epsilon^A+\epsilon^B-\epsilon^C)+2(\lambda_5-\lambda_2)(\epsilon^A-\epsilon^B+\epsilon^C)+2(\lambda_4-\lambda_3)(\epsilon^A-\epsilon^B-\epsilon^C)\\
		&=2(1-\frac{10p}{3}+\frac{32p^2}{9}-\frac{32p^3}{27})(\epsilon^A+\epsilon^B+\epsilon^C)+2(\frac{2p}{3}-\frac{16p^2}{9}+\frac{32p^3}{27})(\epsilon^A+\epsilon^B-\epsilon^C)	.
	\end{split}
\end{equation*}
When $\frac{3}{4}<p\leqslant1$,
\begin{equation*}
	\begin{split}
		\mathcal{C}(\rho_{dep}^{'};H)&=2(\lambda_7-\lambda_0)(\epsilon^A+\epsilon^B+\epsilon^C)+2(\lambda_6-\lambda_1)(\epsilon^A+\epsilon^B-\epsilon^C)+2(\lambda_5-\lambda_2)(\epsilon^A-\epsilon^B+\epsilon^C)+2(\lambda_4-\lambda_3)(\epsilon^A-\epsilon^B-\epsilon^C)\\
		&=2(\frac{2p}{3}-\frac{16p^2}{9}+\frac{32p^3}{27})(\epsilon^A+\epsilon^B+\epsilon^C)+2(1-\frac{10p}{3}+\frac{32p^2}{9}-\frac{32p^3}{27})(\epsilon^A+\epsilon^B-\epsilon^C)	.
	\end{split}
\end{equation*}
So the quantum battery capacity is
\begin{equation*}
	\mathbi{C}(\rho_{dep}^{'},H_{ABC})=\begin{cases}
		2(1-\frac{10p}{3}+\frac{32p^2}{9}-\frac{32p^3}{27})(\epsilon^A+\epsilon^B+\epsilon^C)+2(\frac{2p}{3}-\frac{16p^2}{9}+\frac{32p^3}{27})(\epsilon^A+\epsilon^B-\epsilon^C),  &p\in[0,\frac{3}{4}]\\
		2(\frac{2p}{3}-\frac{16p^2}{9}+\frac{32p^3}{27})(\epsilon^A+\epsilon^B+\epsilon^C)+2(1-\frac{10p}{3}+\frac{32p^2}{9}-\frac{32p^3}{27})(\epsilon^A+\epsilon^B-\epsilon^C),  &p\in(\frac{3}{4},1]
	\end{cases}
\end{equation*}

\subsection{B5. Under the bit-phase flip channel}

After trace-preserving quantum operation, the output state is
	\begin{equation*}
	\begin{split}
		\rho_{bpf}^{'}=\varepsilon(\rho)=E_{000}\rho E_{000}^{\dagger}+E_{001}\rho E_{001}^{\dagger}+E_{010}\rho E_{010}^{\dagger}+E_{100}\rho E_{100}^{\dagger}+E_{011}\rho E_{011}^{\dagger}+E_{101}\rho E_{101}^{\dagger}+E_{110}\rho E_{110}^{\dagger}+E_{111}\rho E_{111}^{\dagger}.
	\end{split}
\end{equation*}
And
	\begin{equation*}
		\begin{split}
			&\rho_{bpf}^{'}=\\
			&\frac{1}{2}\left(\begin{array}{cccccccc}
				1-3p+3p^2 & 0 & 0 & 0 & 0 & 0 & 0 & 1-3p+3p^2-2p^3 \\
				0 & p(1-p) & 0 & 0 & 0 & 0 & -p+3p^2-2p^3 & 0 \\
				0 & 0 & p(1-p) & 0 & 0 & -p+3p^2-2p^3 & 0 & 0 \\
				0 & 0 & 0 & p(1-p) & -p+3p^2-2p^3 & 0 & 0 & 0 \\
				0 & 0 & 0 & -p+3p^2-2p^3 & p(1-p) & 0 & 0 & 0 \\
				0 & 0 & -p+3p^2-2p^3 & 0 & 0 & p(1-p) & 0 & 0 \\
				0 & -p+3p^2-2p^3 & 0 & 0 & 0 & 0 & p(1-p) & 0 \\
				1-3p+3p^2-2p^3 & 0 & 0 & 0 & 0 & 0 & 0 & 1-3p+3p^2
			\end{array}\right)
		\end{split}
	\end{equation*}
The eigenvalues of $\rho_{bpf}^{'}$ are $\lambda_0=p^3$, $\lambda_1=\lambda_2=\lambda_3=p^2-p^3$, $\lambda_4=\lambda_5=\lambda_6=p-2p^2+p^3$ and $\lambda_7=1-3p+3p^2-p^3$. When $0\leqslant p\leqslant\frac{1}{2}$,
\begin{equation*}
	\begin{split}
		\mathcal{C}(\rho_{bpf}^{'};H)&=2(\lambda_7-\lambda_0)(\epsilon^A+\epsilon^B+\epsilon^C)+2(\lambda_6-\lambda_1)(\epsilon^A+\epsilon^B-\epsilon^C)+2(\lambda_5-\lambda_2)(\epsilon^A-\epsilon^B+\epsilon^C)+2(\lambda_4-\lambda_3)(\epsilon^A-\epsilon^B-\epsilon^C)\\
		&=2(1-3p+3p^2-2p^3)(\epsilon^A+\epsilon^B+\epsilon^C)+2(p-3p^2+2p^3)(3\epsilon^A-\epsilon^B-\epsilon^C)	.
	\end{split}
\end{equation*}
When $\frac{1}{2}<p\leqslant1$,
\begin{equation*}
	\begin{split}
		\mathcal{C}(\rho_{bpf}^{'};H)&=2(\lambda_7-\lambda_0)(\epsilon^A+\epsilon^B+\epsilon^C)+2(\lambda_6-\lambda_1)(\epsilon^A+\epsilon^B-\epsilon^C)+2(\lambda_5-\lambda_2)(\epsilon^A-\epsilon^B+\epsilon^C)+2(\lambda_4-\lambda_3)(\epsilon^A-\epsilon^B-\epsilon^C)\\
		&=2(-1+3p-3p^2+2p^3)(\epsilon^A+\epsilon^B+\epsilon^C)+2(-p+3p^2-2p^3)(3\epsilon^A-\epsilon^B-\epsilon^C)	.
	\end{split}
\end{equation*}
Therefore, the quantum battery capacity of the output state $\rho_{bpf}^{'}$ is
\begin{equation*}
	\mathbi{C}(\rho_{bpf}^{'},H_{ABC})=\begin{cases}
		2(1-3p+3p^2-2p^3)(\epsilon^A+\epsilon^B+\epsilon^C)+2(p-3p^2+2p^3)(3\epsilon^A-\epsilon^B-\epsilon^C),&p\in[0,\frac{1}{2}]\\
		2(-1+3p-3p^3+2p^3)(\epsilon^A+\epsilon^B+\epsilon^C)+2(-p+3p^2-2p^3)(3\epsilon^A-\epsilon^B-\epsilon^C),&p\in(\frac{1}{2},1]
	\end{cases}
\end{equation*}

\subsection{B6. Under the amplitude damping channel}

The output state of GHZ state on the first subsystem is	
	$$
	\rho_{adc}^{'}=(E_0 \otimes I \otimes I)\rho(E_{0}^{\dagger} \otimes I \otimes I)+(E_1 \otimes I \otimes I)\rho(E_{1}^{\dagger} \otimes I \otimes I),
	$$
	and
	$$ 
	\rho_{adc}^{'}=\frac{1}{2}\left(\begin{array}{cccccccc}
		1 & 0 & 0 & 0 & 0 & 0 & 0 & \sqrt{1-p}\\
		0 & 0 & 0 & 0 & 0 & 0 & 0 & 0\\
		0 & 0 & 0 & 0 & 0 & 0 & 0 & 0\\
		0 & 0 & 0 & p & 0 & 0 & 0 & 0\\
		0 & 0 & 0 & 0 & 0 & 0 & 0 & 0\\
		0 & 0 & 0 & 0 & 0 & 0 & 0 & 0\\
		0 & 0 & 0 & 0 & 0 & 0 & 0 & 0\\
		\sqrt{1-p} & 0 & 0 & 0 & 0 & 0 & 0 & 1-p
	\end{array}\right).
	$$\\
	So the eigenvalues of $\rho_{adc}^{'}$ are $\lambda_i=0$, (i=0,...,5), $\lambda_6=\frac{p}{2}$, $\lambda_7=1-\frac{p}{2}$. Therefore, the order of eigenvalues is $\lambda_0=\lambda_1=...=\lambda_5\leqslant \lambda_6\leqslant\lambda_7$. Set $\epsilon^A=0.5$, $\epsilon^B=0.3$, and $\epsilon^C=0.1$. Using (\ref{e1}), we have 
	\begin{equation*}
		\begin{split}
			\mathcal{C}(\rho_{bpf}^{'};H)&=2(\lambda_7-\lambda_0)(\epsilon^A+\epsilon^B+\epsilon^C)+2(\lambda_6-\lambda_1)(\epsilon^A+\epsilon^B-\epsilon^C)+2(\lambda_5-\lambda_2)(\epsilon^A-\epsilon^B+\epsilon^C)+2(\lambda_4-\lambda_3)(\epsilon^A-\epsilon^B-\epsilon^C)\\
			&=2(1-\frac{p}{2})(\epsilon^A+\epsilon^B+\epsilon^C)+p(\epsilon^A+\epsilon^B-\epsilon^C)	.
		\end{split}
		\end{equation*}
	
\section{Appendix C: Quantum battery capacity of GHZ state under n times Markovian channels.}

\subsection{C1. Under the n times phase flip channel}

The output state after n times phase flip channels is
\begin{equation*}
	\begin{split}
		\rho_{pf}^{n}=\frac{1}{2}\left(\begin{array}{cccccccc}
			1 & 0 & 0 & 0 & 0 & 0 & 0 & (1-2p)^{3n} \\
			0 & 0 & 0 & 0 & 0 & 0 & 0 & 0 \\
			0 & 0 & 0 & 0 & 0 & 0 & 0 & 0 \\
			0 & 0 & 0 & 0 & 0 & 0 & 0 & 0 \\
			0 & 0 & 0 & 0 & 0 & 0 & 0 & 0 \\
			0 & 0 & 0 & 0 & 0 & 0 & 0 & 0 \\
			0 & 0 & 0 & 0 & 0 & 0 & 0 & 0 \\
			(1-2p)^{3n} & 0 & 0 & 0 & 0 & 0 & 0 & 1
		\end{array}\right)
	\end{split}
\end{equation*}
The eigenvalues of $\rho_{pf}^{n}$ are $\lambda_i=0,(i=0,...,5)$, $\lambda_6=\frac{1}{2}-\frac{\sqrt{(1-2p)^{6n}}}{2}$ and $\lambda_7=\frac{1}{2}+\frac{\sqrt{(1-2p)^{6n}}}{2}$. Therefore,
\begin{equation*}
	\begin{split}
		\mathcal{C}(\rho_{pf}^{n};H)&=2(\lambda_7-\lambda_0)(\epsilon^A+\epsilon^B+\epsilon^C)+2(\lambda_6-\lambda_1)(\epsilon^A+\epsilon^B-\epsilon^C)+2(\lambda_5-\lambda_2)(\epsilon^A-\epsilon^B+\epsilon^C)+2(\lambda_4-\lambda_3)(\epsilon^A-\epsilon^B-\epsilon^C)\\
		&=(1+\sqrt{(1-2p)^{6n}})(\epsilon^A+\epsilon^B+\epsilon^C)+(1-\sqrt{(1-2p)^{6n}})(\epsilon^A+\epsilon^B-\epsilon^C)	.
	\end{split}
\end{equation*}

\subsection{C2. Under the n times dephasing channel}

The output state after n times dephasing channels is
\begin{equation*}
	\begin{split}
		\rho_{dp}^{n}=\frac{1}{2}\left(\begin{array}{cccccccc}
			1 & 0 & 0 & 0 & 0 & 0 & 0 & (1-p)^{3n} \\
			0 & 0 & 0 & 0 & 0 & 0 & 0 & 0 \\
			0 & 0 & 0 & 0 & 0 & 0 & 0 & 0 \\
			0 & 0 & 0 & 0 & 0 & 0 & 0 & 0 \\
			0 & 0 & 0 & 0 & 0 & 0 & 0 & 0 \\
			0 & 0 & 0 & 0 & 0 & 0 & 0 & 0 \\
			0 & 0 & 0 & 0 & 0 & 0 & 0 & 0 \\
			(1-p)^{3n} & 0 & 0 & 0 & 0 & 0 & 0 & 1
		\end{array}\right)
	\end{split}
\end{equation*}
Therefore, the eigenvalues of $\rho_{dp}^{n}$ are $\lambda_i=0,(i=0,...,5)$, $\lambda_6=\frac{1}{2}-\frac{\sqrt{(1-p)^{6n}}}{2}$ and $\lambda_7=\frac{1}{2}+\frac{\sqrt{(1-p)^{6n}}}{2}$.
So,
\begin{equation*}
	\begin{split}
		\mathcal{C}(\rho_{dp}^{n};H)&=2(\lambda_7-\lambda_0)(\epsilon^A+\epsilon^B+\epsilon^C)+2(\lambda_6-\lambda_1)(\epsilon^A+\epsilon^B-\epsilon^C)+2(\lambda_5-\lambda_2)(\epsilon^A-\epsilon^B+\epsilon^C)+2(\lambda_4-\lambda_3)(\epsilon^A-\epsilon^B-\epsilon^C)\\
		&=(1+\sqrt{(1-p)^{6n}})(\epsilon^A+\epsilon^B+\epsilon^C)+(1-\sqrt{(1-p)^{6n}})(\epsilon^A+\epsilon^B-\epsilon^C)	.
	\end{split}
\end{equation*}

\subsection{C3. Under the  n times amplitude damping channel}

After going through the amplitude damping channel n times on the first subsystem, the output state $\rho_{adc}^{(n)}$ is the 
	\begin{equation*}
		\begin{split}
			\rho_{adc}^{(n)}=(E_0\otimes I\otimes I)\rho_{adc}^{(n-1)}(E_0^{\dagger}\otimes I\otimes I)+E_1\otimes I\otimes I)\rho_{adc}^{(n-1)}(E_1^{\dagger}\otimes I\otimes I),
		\end{split}
	\end{equation*}
	which can be rewritten as
	$$ \rho_{adc}^{(n)}=\sum_{i_1,1_2,...,i_n=0,1}(E_{i_1i_2...i_n}\otimes I\otimes I)\rho_{ABC}(E_{i_1i_2...i_n}^{\dagger}\otimes I\otimes I)
	$$ 
	with $E_{i_1i_2...i_n}=E_{i_1}E_{i_2}...E_{i_n}$. 
	Utilizing the properties of operators $E_0$ and $E_1$ in the amplitude damping channel, $E_{1}^{2}=0$, $E_0E_1=E_1$ and $E_1E_0=\sqrt{1-p}E_1$, 
	$\rho_{adc}^{(n)}$ is simplified to be 
	\begin{equation*}
		\begin{split}
			\rho_{adc}^{(n)}&=(E_{0}^{n}\otimes I\otimes I)\rho_{ABC}((E_{0}^{n})^{\dagger}\otimes I\otimes I)+\sum_{i=0}^{n-1}(E_1E_{0}^{n-i-1}\otimes I\otimes I)\rho_{ABC}((E_1E_{0}^{n-i-1})^{\dagger}\otimes I\otimes I)\\
			&=(E_{0}^{n}\otimes I\otimes I)\rho_{ABC}((E_{0}^{n})^{\dagger}\otimes I\otimes I)+(E_1E_{0}^{n-2}\otimes I \otimes I)\rho_{ABC}((E_1E_{0}^{n-2})^{\dagger}\otimes I \otimes I)\\
			&+(E_1E_{0}^{n-2}\otimes I \otimes I)\rho_{ABC}((E_1E_{0}^{n-2})^{\dagger}\otimes I \otimes I)+...+(E_1E_0\otimes I \otimes I)\rho_{ABC}((E_1E_0)^{\dagger}\otimes I \otimes I),  		
		\end{split}
	\end{equation*}
	namely,
	$$
	\rho_{adc}^{n}=\frac{1}{2}\left(\begin{array}{cccccccc}
		1 & 0 & 0 & 0 & 0 & 0 & 0 & (1-p)^{\frac{n}{2}}\\
		0 & 0 & 0 & 0 & 0 & 0 & 0 & 0\\
		0 & 0 & 0 & 0 & 0 & 0 & 0 & 0\\
		0 & 0 & 0 & 1-(1-p)^n & 0 & 0 & 0 & 0\\
		0 & 0 & 0 & 0 & 0 & 0 & 0 & 0\\
		0 & 0 & 0 & 0 & 0 & 0 & 0 & 0\\
		0 & 0 & 0 & 0 & 0 & 0 & 0 & 0\\
		(1-p)^{\frac{n}{2}} & 0 & 0 & 0 & 0 & 0 & 0 & (1-p)^n
	\end{array}\right).
	$$
	The eigenvalues of $\rho_{adc}^{(n)}$ are $\lambda_{i}^{(n)}=0, i=(0,1,2,3,4,5)$, $\lambda_{6}^{(n)}=\frac{1-\sqrt{(1-p)^{2n}}}{2}$, $\lambda_{7}^{(n)}= \frac{1+\sqrt{(1-p)^{2n}}}{2}$. We always have $\lambda_{i}^{(n)}<\lambda_{6}^{(n)}<\lambda_{7}^{(n)}$ for n=2, 3, 4, 10, 100. Set $\epsilon^A=0.5$, $\epsilon^B=0.3$, $\epsilon^C=0.1$. From (\ref{e3}), we have
	\begin{equation*}
		\begin{split}
			\mathbi{C}(\rho_{adc}^{(n)},H_{ABC})&=2(\lambda_7-\lambda_0)(\epsilon^A+\epsilon^B+\epsilon^C)+2(\lambda_6-\lambda_1)(\epsilon^A+\epsilon^B-\epsilon^C)\\
			&=(1+\sqrt{(1-p)^{2n}})\times 0.9+(1-\sqrt{(1-p)^{2n}})\times 0.7\\
			&=0.2\sqrt{(1-p)^{2n}}+1.6.
		\end{split}
\end{equation*}

\section{Appendix D: Quantum battery capacity of GHZ state under tri-side phase flip channel of the same type.}
\subsection{D1. Under the tri-side same type phase flip channel}

Under the tri-side phase flip channel of the same type, the output state if GHZ state is
	\begin{equation*}
	\begin{split}
		\rho_{pf-pf-pf}^{'}&=\varepsilon(\rho)\\
		&=(E_0^{A}\otimes E_0^{B} \otimes E_0^{C})\rho (E_0^{A}\otimes E_0^{B} \otimes E_0^{C})^{\dagger}+(E_0^{A}\otimes E_0^{B} \otimes E_1^{C})\rho (E_0^{A}\otimes E_0^{B} \otimes E_1^{C})^{\dagger}\\
		&+(E_0^{A}\otimes E_1^{B} \otimes E_0^{C})\rho (E_0^{A}\otimes E_1^{B} \otimes E_0^{C})^{\dagger}+(E_1^{A}\otimes E_0^{B} \otimes E_0^{C})\rho (E_1^{A}\otimes E_0^{B} \otimes E_0^{C})^{\dagger}\\
		&+(E_0^{A}\otimes E_1^{B} \otimes E_1^{C})\rho (E_0^{A}\otimes E_1^{B} \otimes E_1^{C})^{\dagger}+(E_1^{A}\otimes E_0^{B} \otimes E_1^{C})\rho (E_1^{A}\otimes E_0^{B} \otimes E_1^{C})^{\dagger}\\
		&+(E_1^{A}\otimes E_1^{B} \otimes E_0^{C})\rho (E_1^{A}\otimes E_1^{B} \otimes E_0^{C})^{\dagger}+(E_1^{A}\otimes E_1^{B} \otimes E_1^{C})\rho (E_1^{A}\otimes E_1^{B} \otimes E_1^{C})^{\dagger}.
	\end{split}
\end{equation*}
So the eigenvalues of $\rho_{pf-pf-pf}^{'}$ are $\lambda_i=0,(i=0,...,5)$, $\lambda_6=\frac{1-(1-2p)(1-2q)(1-2\gamma)}{2}$ and $\lambda_7=\frac{1+(1-2p)(1-2q)(1-2\gamma)}{2}$. We have
\begin{equation*}
	\begin{split}
		\mathcal{C}(\rho_{pf-pf-pf}^{'};H)&=2(\lambda_7-\lambda_0)(\epsilon^A+\epsilon^B+\epsilon^C)+2(\lambda_6-\lambda_1)(\epsilon^A+\epsilon^B-\epsilon^C)\\
		&+2(\lambda_5-\lambda_2)(\epsilon^A-\epsilon^B+\epsilon^C)+2(\lambda_4-\lambda_3)(\epsilon^A-\epsilon^B-\epsilon^C)\\
		&=(1+(1-2p)(1-2q)(1-2\gamma))(\epsilon^A+\epsilon^B+\epsilon^C)+(1-(1-2p)(1-2q)(1-2\gamma)(\epsilon^A+\epsilon^B-\epsilon^C)	
	\end{split}
\end{equation*}
We assume that $\gamma$ is set to $0.25, 0.5, 0.75$ and $1$. Specifically, when $\gamma=0.5$, the battery capacity remains a fixed value and does not vary with changes in $p$ and $q$.

\subsection{D2. Under the n times tri-side phase flip channel of same type}

Similarly, the eigenvalues of $\rho_{pf-pf-pf}^{n}$ are $\lambda_i=0,(i=0,..,5)$, $\lambda_6=\frac{1}{2}-\frac{\sqrt{(1-2p)^{2n}(1-2q)^{2n}(1-2\gamma)^{2n}}}{2}$ and $\lambda_7=\frac{1}{2}+\frac{\sqrt{(1-2p)^{2n}(1-2q)^{2n}(1-2\gamma)^{2n}}}{2}$. Therefore, according to (\ref{e1}), we have
\begin{equation*}
	\begin{split}
		\mathcal{C}(\rho_{pf-pf-pf}^{n};H)&=2(\lambda_7-\lambda_0)(\epsilon^A+\epsilon^B+\epsilon^C)+2(\lambda_6-\lambda_1)(\epsilon^A+\epsilon^B-\epsilon^C)\\
		&+2(\lambda_5-\lambda_2)(\epsilon^A-\epsilon^B+\epsilon^C)+2(\lambda_4-\lambda_3)(\epsilon^A-\epsilon^B-\epsilon^C)\\
		&=(1+\sqrt{(1-2p)^{2n}(1-2q)^{2n}(1-2\gamma)^{2n}})(\epsilon^A+\epsilon^B+\epsilon^C)\\
		&+(1-\sqrt{(1-2p)^{2n}(1-2q)^{2n}(1-2\gamma)^{2n}})(\epsilon^A+\epsilon^B-\epsilon^C)	.
	\end{split}
\end{equation*}

\section{Appendix E: Quantum battery capacity of GHZ-like state}
\subsection{E1. Under the phase flip channel}

Under the phase flip channel, the output state of GHZ-like state is
$$
\rho_{pf}^{'}=\left(\begin{array}{cccccccc}
	c_1 & 0 & 0 & 0 & 0 & 0 & 0 & c_2(1-2p)^3\\
	0 & 0 & 0 & 0 & 0 & 0 & 0 & 0\\
	0 & 0 & 0 & 0 & 0 & 0 & 0 & 0\\
	0 & 0 & 0 & 0 & 0 & 0 & 0 & 0\\
	0 & 0 & 0 & 0 & 0 & 0 & 0 & 0\\
	0 & 0 & 0 & 0 & 0 & 0 & 0 & 0\\
	0 & 0 & 0 & 0 & 0 & 0 & 0 & 0\\
	c_2(1-2p)^3 & 0 & 0 & 0 & 0 & 0 & 0 & c_3
\end{array}\right).
$$
The eigenvalues of $\rho_{pf}^{'}$ are $\lambda_i=0,(i=0,...,5)$, $\lambda_6=\frac{c_1+c_3-\sqrt{(c_1-c_3)^2+4c_{2}^{2}(1-2p)^6}}{2}$ and $\lambda_6=\frac{c_1+c_3+\sqrt{(c_1-c_3)^2+4c_{2}^{2}(1-2p)^6}}{2}$
And
\begin{equation*}
	\begin{split}
		\mathcal{C}(\rho_{pf}^{'};H)=(1+\sqrt{1-4c_2c_3(1-(1-2p)^6)})(\epsilon^A+\epsilon^B+\epsilon^C)+(1-\sqrt{1-4c_2c_3(1-(1-2p)^6)})(\epsilon^A+\epsilon^B-\epsilon^C)	.
	\end{split}
\end{equation*}
Similarly,under the n times phase flip channel, the output state of GHZ-like states is
$$
\rho_{pf}^{n}=\left(\begin{array}{cccccccc}
	c_1 & 0 & 0 & 0 & 0 & 0 & 0 & c_2(1-2p)^{3n}\\
	0 & 0 & 0 & 0 & 0 & 0 & 0 & 0\\
	0 & 0 & 0 & 0 & 0 & 0 & 0 & 0\\
	0 & 0 & 0 & 0 & 0 & 0 & 0 & 0\\
	0 & 0 & 0 & 0 & 0 & 0 & 0 & 0\\
	0 & 0 & 0 & 0 & 0 & 0 & 0 & 0\\
	0 & 0 & 0 & 0 & 0 & 0 & 0 & 0\\
	c_2(1-2p)^{3n} & 0 & 0 & 0 & 0 & 0 & 0 & c_3
\end{array}\right).
$$
The eigenvalues of $\rho_{pf}^{n}$ are $\lambda_i=0,(i=0,...,5)$, $\lambda_6=\frac{c_1+c_3-\sqrt{(c_1-c_3)^2+4c_{2}^{2}(1-2p)^{6n}}}{2}$ and $\lambda_6=\frac{c_1+c_3+\sqrt{(c_1-c_3)^2+4c_{2}^{2}(1-2p)^{6n}}}{2}$.
And
\begin{equation*}
	\begin{split}
		\mathcal{C}(\rho_{pf}^{n};H)&=(c_1+c_3+\sqrt{(c_1-c_3)^2+4c_{2}^{2}(1-2p)^{6n}})(\epsilon^A+\epsilon^B+\epsilon^C)\\
		&+(c_1+c_3-\sqrt{(c_1-c_3)^2+4c_{2}^{2}(1-2p)^{6n}})(\epsilon^A+\epsilon^B-\epsilon^C)	.
	\end{split}
\end{equation*}
Likewise, under the tri-side phase flip channel of same type, the capacity of output state is
	\begin{equation*}
		\begin{split}
			\mathcal{C}(\rho_{pf-pf-pf}^{'};H)&=(c_1+c_3+\sqrt{(c_1-c_3)^2+4c_{2}^{2}(1-2p)^2(1-2q)^2(1-2\gamma)^2})(\epsilon^A+\epsilon^B+\epsilon^C)\\
			&+(c_1+c_3-\sqrt{(c_1-c_3)^2+4c_{2}^{2}(1-2p)^2(1-2q)^2(1-2\gamma)^2})(\epsilon^A+\epsilon^B-\epsilon^C)	.
		\end{split}
	\end{equation*}
And under the n times tri-side phase flip channel of same type, the capacity of output state is
	\begin{equation*}
		\begin{split}
			\mathcal{C}(\rho_{pf-pf-pf}^{n};H)&=(c_1+c_3+\sqrt{(c_1-c_3)^2+4c_{2}^{2}(1-2p)^{2n}(1-2q)^{2n}(1-2\gamma)^{2n}})(\epsilon^A+\epsilon^B+\epsilon^C)\\
			&+(c_1+c_3-\sqrt{(c_1-c_3)^2+4c_{2}^{2}(1-2p)^{2n}(1-2q)^{2n}(1-2\gamma)^{2n}})(\epsilon^A+\epsilon^B-\epsilon^C)	.
		\end{split}
\end{equation*}
	
	\subsection{E2. Under the dephasing channel}

Under the dephasing channel, the output state of GHZ-like state is
	$$
	\rho_{dp}^{'}=\left(\begin{array}{cccccccc}
		c_1 & 0 & 0 & 0 & 0 & 0 & 0 & c_2(1-p)^3\\
		0 & 0 & 0 & 0 & 0 & 0 & 0 & 0\\
		0 & 0 & 0 & 0 & 0 & 0 & 0 & 0\\
		0 & 0 & 0 & 0 & 0 & 0 & 0 & 0\\
		0 & 0 & 0 & 0 & 0 & 0 & 0 & 0\\
		0 & 0 & 0 & 0 & 0 & 0 & 0 & 0\\
		0 & 0 & 0 & 0 & 0 & 0 & 0 & 0\\
		c_2(1-p)^3 & 0 & 0 & 0 & 0 & 0 & 0 & c_3
	\end{array}\right).
	$$
	The eigenvalues of $\rho_{pf}^{'}$ are $\lambda_i=0,(i=0,...,5)$, $\lambda_6=\frac{c_1+c_3-\sqrt{(c_1-c_3)^2+4c_{2}^{2}(1-p)^6}}{2}$ and $\lambda_6=\frac{c_1+c_3+\sqrt{(c_1-c_3)^2+4c_{2}^{2}(1-p)^6}}{2}$
	And
	\begin{equation*}
		\begin{split}
			\mathcal{C}(\rho_{dp}^{'};H)=(1+\sqrt{1-4c_2c_3(1-(1-p)^6)})(\epsilon^A+\epsilon^B+\epsilon^C)+(1-\sqrt{1-4c_2c_3(1-(1-p)^6)})(\epsilon^A+\epsilon^B-\epsilon^C)	.
		\end{split}
	\end{equation*}
	And under the n times dephasing channel, the output state of GHZ-like state is
	$$
	\rho_{dp}^{n}=\left(\begin{array}{cccccccc}
		c_1 & 0 & 0 & 0 & 0 & 0 & 0 & c_2(1-p)^{3n}\\
		0 & 0 & 0 & 0 & 0 & 0 & 0 & 0\\
		0 & 0 & 0 & 0 & 0 & 0 & 0 & 0\\
		0 & 0 & 0 & 0 & 0 & 0 & 0 & 0\\
		0 & 0 & 0 & 0 & 0 & 0 & 0 & 0\\
		0 & 0 & 0 & 0 & 0 & 0 & 0 & 0\\
		0 & 0 & 0 & 0 & 0 & 0 & 0 & 0\\
		c_2(1-p)^{3n} & 0 & 0 & 0 & 0 & 0 & 0 & c_3
	\end{array}\right).
	$$
	The eigenvalues of $\rho_{pf}^{n}$ are $\lambda_i=0,(i=0,...,5)$, $\lambda_6=\frac{c_1+c_3-\sqrt{(c_1-c_3)^2+4c_{2}^{2}(1-p)^{6n}}}{2}$ and $\lambda_6=\frac{c_1+c_3+\sqrt{(c_1-c_3)^2+4c_{2}^{2}(1-p)^{6n}}}{2}$.
	And the quantum battery capacity is 
	\begin{equation*}
		\begin{split}
			\mathcal{C}(\rho_{dp}^{n};H)=(c_1+c_3+\sqrt{(c_1-c_3)^2+4c_{2}^{2}(1-p)^{6n}})(\epsilon^A+\epsilon^B+\epsilon^C)+(c_1+c_3-\sqrt{(c_1-c_3)^2+4c_{2}^{2}(1-p)^{6n}})(\epsilon^A+\epsilon^B-\epsilon^C)	.
		\end{split}
	\end{equation*}

\subsection{E3. Under the n times amplitude damping channel}

Under going through the amplitude damping channel n times on the first subsystem, the output state $\rho_{adc}^{(n)}$ is the 
\begin{equation*}
	\begin{split}
		\rho_{adc}^{(n)}=(E_{0}^{n}\otimes I\otimes I)\rho_{ABC}((E_{0}^{n})^{\dagger}\otimes I\otimes I)+\sum_{i=0}^{n-1}(E_1E_{0}^{n-i-1}\otimes I\otimes I)\rho_{ABC}((E_1E_{0}^{n-i-1})^{\dagger}\otimes I\otimes I),
	\end{split}
\end{equation*}
namely,
\begin{equation*}
	\begin{split}
		\rho_{adc}^{n}=\left(\begin{array}{cccccccc}
			c_1 & 0 & 0 & 0 & 0 & 0 & 0 & c_2(1-p)^{\frac{n}{2}}\\
			0 & 0 & 0 & 0 & 0 & 0 & 0 & 0\\
			0 & 0 & 0 & 0 & 0 & 0 & 0 & 0\\
			0 & 0 & 0 & c_3[1-(1-p)^n] & 0 & 0 & 0 & 0\\
			0 & 0 & 0 & 0 & 0 & 0 & 0 & 0\\
			0 & 0 & 0 & 0 & 0 & 0 & 0 & 0\\
			0 & 0 & 0 & 0 & 0 & 0 & 0 & 0\\
			c_2(1-p)^{\frac{n}{2}} & 0 & 0 & 0 & 0 & 0 & 0 & c_3(1-p)^n
		\end{array}\right).
	\end{split}
\end{equation*}
The eigenvalues of $\rho_{adc}^{(n)}$ are $\lambda_{i}=0, i=(0,1,2,3,4)$, $\lambda_{5}=c_3[1-(1-p)^n]$, $\lambda_{6}=\frac{c_1+c_3(1-p)^n-\sqrt{(c_1-c_3(1-p)^{n})^2+4c_{2}^{2}(1-p)^n}}{2}$,
$\lambda_{7}= \frac{c_1+c_3(1-p)^n+\sqrt{(c_1-c_3(1-p)^{n})^2+4c_{2}^{2}(1-p)^n}}{2}$. Set $\epsilon^A=0.5$, $\epsilon^B=0.3$, $\epsilon^C=0.1$. Using (\ref{e1}), when $c_1=a^2=0.25$, we have
\begin{equation*}
	\begin{split}
		\mathbi{C}(\rho_{adc}^{n},H_{ABC})=\begin{cases}
			2\lambda_7(\epsilon^A+\epsilon^B+\epsilon^C)+2\lambda_5(\epsilon^A+\epsilon^B-\epsilon^C),  &p\in[0,x]\\
			2\lambda_5(\epsilon^A+\epsilon^B+\epsilon^C)+2\lambda_7(\epsilon^A+\epsilon^B-\epsilon^C),  &p\in(x,1],
		\end{cases}
	\end{split}
\end{equation*}
where, x is the intersection point of the eigenvalues $\lambda_5$ and $\lambda_7$. When $c_1=a^2=0.75$, we have
\begin{equation*}
\mathbi{C}(\rho_{adc}^{n},H_{ABC})
=2\lambda_7(\epsilon^A+\epsilon^B+\epsilon^C)+2\lambda_5(\epsilon^A+\epsilon^B-\epsilon^C).
\end{equation*}

\appendix

\end{document}